\documentclass[a4paper,11pt]{article}
\usepackage{pstricks, pst-node, pst-plot, pst-coil}
\usepackage{mathrsfs, amsmath, amsthm, amssymb, a4, epsfig,
appendix}
\usepackage{citesort}
\usepackage{mcite}
\usepackage{axodraw}
\usepackage{epsf}
\renewcommand{\d}{\textup{d}}
\newcommand{\e}{\textup{e}}
\newpsobject{showgrid}{psgrid}{subgriddiv=1,griddots=10,gridlabels=6pt}

\title{Real fermionic symmetry in type II supergravity}

\begin{document}

\begin{titlepage}
\hfill DAMTP-2010-61
\null
\vspace{0.75cm}
\begin{center}

{{\LARGE  \bf Real Fermionic Symmetry in Type II Supergravity}} \\

\vskip 1.5cm {Hadi Godazgar $^\dag$\footnote{H.M.Godazgar@damtp.cam.ac.uk} and Malcolm J. Perry$^{\dag,\ddag}$\footnote{M.J.Perry@damtp.cam.ac.uk}}
\\
{\vskip 0.2cm
$^\dag$DAMTP, Centre for Mathematical Science,\\
University of Cambridge,\\
Wilberforce Road, Cambridge CB3 0WA, England\\
}
{\vskip 0.2cm
$^\ddag$Perimeter Institute for Theoretical Physics,\\
31 Caroline St. N., Waterloo, \\
Ontario N2L 2Y5, Canada}
\end{center}
\vskip 1 cm

\begin{center}
 \today
\end{center}

\vskip 1cm

\begin{abstract}
\baselineskip=18pt\
It is known that the transformations of fermionic T-duality, derived from the worldsheet theory, generically transform real supergravity backgrounds to complex supergravity backgrounds. We consider the low-energy target space theory and show that the type II supergravity equations admit a symmetry that transforms the Ramond-Ramond fields and the dilaton. The transformations given by this symmetry involve Killing spinors and include the transformations of Berkovits and Maldacena. However, we show that they also allow real transformations.
\end{abstract}

\end{titlepage}

\section{Introduction}

In the past two decades, T-duality \cite{giveon, lectdual} has been highly successful in helping increase our understanding of string theory. In the 1990s, S, T and U-dualities were used to relate the five different string theories. The existence of these dualities was crucial in the conjecture that the five string theories are different limits of a theory that is the strong coupling limit of type IIA string theory, M-theory \cite{mtheoryht, mtheoryw}. T-duality has also provided insights into the study of D-branes, which would perhaps be inaccessible otherwise. For example, Myers \cite{myerseffect} used consistency of the world-volume action for the D-brane with T-duality to find the correct coupling of background Ramond-Ramond fields to D-branes, which was used to discover the Myers effect. Furthermore, another important application of T-duality has been to use it to generate solutions in supergravity \cite{Tsolngen}.

In Buscher's formulation of T-duality \cite{buscher1, *buscher2}, a shift symmetry of a target space coordinate, which corresponds to an isometry in the target space of the sigma-model, is used to make a field redefinition in the sigma model. The new sigma model is classically of the same form as the original sigma model except for the sigma model couplings, i.e.\ the metric and the 2-form field, which are different. The two sigma models are equivalent quantum-mechanically if the dilaton also transforms. This shows that the string theories described by the two sigma models with different couplings, which correspond to different backgrounds for the string theories, are equivalent. The transformed background fields are related to the original fields by the killing vector that corresponds to the isometry. 

Recently, this idea has been generalised to the case where the target space has a fermionic isometry, or supersymmetry, as opposed to an isometry, to find a duality of tree-level type II string theory, fermionic T-duality \cite{fermdual, fermdual2}. Under this duality the background Ramond-Ramond fields and the dilaton transform and the metric and the NSNS 2-form field are invariant. Analogously to T-duality, the transformation of the background supergravity fields are given by the Killing spinors corresponding to the supersymmetry in superspace.

Given the success of T-duality, we expect that fermionic T-duality will also make important contributions to our understanding of string theory. In fact, fermionic T-duality was introduced to explain the dual superconformal symmetry of planar scattering amplitudes in $\mathcal{N}=4$ super Yang-Mills theory \cite{fermdual, fermdual2}, which has no obvious origin in the weak coupling computations of these amplitudes in which this symmetry was found \cite{dualsuperconf1, *dualsuperconf2}. There have also been other studies of fermionic T-duality \cite{adam, fre, hao, bb, sfetsos}.

However, in contrast to T-duality, fermionic T-duality generically transforms a real supergravity background into a complex supergravity background. An ordinary T-duality along a time-like direction can then be applied to get back a real background. This means that the application of fermionic T-duality as a solution generating mechanism, one of the key applications of T-duality, is limited to supergravity solutions with a timelike Killing vector. 

In this paper, we consider fermionic T-duality from the spacetime viewpoint, rather than the worldsheet perspective in which it was found. We consider a general ansatz for the transformation of the Ramond-Ramond fields and the dilaton involving Killing spinors in both type IIA and type IIB supergravity. We then systematically impose that the supergravity equations are invariant under this transformation, i.e.\ we impose that the transformed fields are solutions of the supergravity equations. We find that the symmetry includes fermionic T-duality, as it must, but it also admits real transformations of the supergravity fields.

The structure of the paper is as follows. In section \ref{review}, we review B\"{u}scher's T-duality and show how a symmetry in the target space allows a field redefinition in the worldsheet theory that gives rise to a duality. We also review the transformation rules of fermionic T-duality \cite{fermdual} in this section. Then, in section \ref{symmIIA}, we set our conventions by stating the type IIA supergravity Lagrangian and equations, and we construct the symmetry for type IIA supergravity. In section \ref{symmIIB}, we construct an analogous symmetry for type IIB supergravity. Finally, in section \ref{com}, we make some comments and outline future work.

\section{Review of T-duality and fermionic T-duality}
\label{review}

B\"{u}scher \cite{buscher1, *buscher2} showed that T-duality in curved backgrounds arises as a symmetry of the sigma-model. Consider the bosonic string sigma-model 
\begin{equation}
S= \frac{1}{4 \pi \alpha'} \int \d ^2 {\sigma} \left[ \sqrt{h} h^{ \alpha \beta} g_{a b} \partial_{\alpha} X^{a}\partial_{\beta} X^{b}+ \epsilon^{\alpha \beta} B_{a b} \partial_{\alpha} X^{a}\partial_{\beta} X^{b} + \alpha' \sqrt{h} R^{(2)} \phi (X) \right]. \notag
\end{equation}
The field $X^{a}$ is the position of the point $(\sigma^{1},\sigma^{2})$ on the worldsheet in spacetime; $g_{a b}$ is the metric on the target space; $B_{a b}$ is the antisymmetric gauge potential; $\phi$ is the dilaton and $R^{(2)}$ is the curvature of the worldsheet metric $h.$ Imposing conformal invariance in the quantum theory gives the equations of motion for the background fields \cite{stringback}. In 26 dimensions, the equations of motion for the metric, two-form field and dilaton are
\begin{gather}
R_{a b}-\frac{1}{4} H_{a}^{\;\; c d} H_{b c d} +2 \nabla_{a} \nabla_{b} \phi=0, \notag \\
\nabla_{c} H^{c}_{\;\;a b} -2 \left(\partial_{c} \phi \right) H^{c}_{\;\;ab}=0, \notag \\
4 \left(\partial \phi \right)^2 -4 \Box \phi -R + \frac{1}{12} H^2=0, 
\label{betaeqns}
\end{gather}
respectively. The tensor $R_{a b}$ is the Ricci tensor associated to the metric on the target space and $H_{a b c} = 3 \partial_{[a} B_{b c]}.$

If there is a Killing vector in the target space, $k,$ then we can choose a coordinate system---we will let $X^{a}$ be such a coordinate system---in which $k=\partial/\partial{X^0}.$ In this coordinate system, the metric, two-form field and the dilaton are independent of the $X^0$ coordinate. We can then write $$ \partial_{\alpha} X^0 =V_{\alpha}, $$ in the action, but we must impose the constraint that $V_{\alpha}$ is exact. For Euclidean worldsheets of spherical topology we can impose this constraint using a Lagrange multiplier term $$ \epsilon^{\alpha \beta} \hat{X}^{0} \partial_{\alpha} V_{\beta}.$$ 

So we can write the action as 
\begin{multline}
S=\frac{1}{4 \pi \alpha'} \int \d ^2 {\sigma} \left( \sqrt{h} h^{ \alpha \beta} \left[ g_{0 0} V_{\alpha} V_{\beta}  + 2 g_{0 i} V_{\alpha} \partial_{\beta} X^{i} + g_{i j} \partial_{\alpha} X^{i} \partial_{\beta} X^{j} \right] \right. \\ \left. +  \epsilon^{\alpha \beta} \left[2 B_{0 i} V_{\alpha} \partial_{\beta} X^{i} +B_{i j} \partial_{\alpha}X^{i} \partial_{\beta} X^{j} \right] + 2  \epsilon^{\alpha \beta} \hat{X}^{0} \partial_{\alpha} V_{\beta} + \alpha' \sqrt{h} R^{(2)} \phi (X) \right), \notag
\end{multline}
where $a=(0,i).$ 

The equation of motion for $\hat{X}^{0}$ gives that $V$ is closed, which for a spherical worldsheet implies that $V$ is exact, so we get back the original theory. The $V$ equation of motion is $$ V_{\alpha} =-\frac{1}{g_{00}} \left[ g_{0i} \partial_{\alpha} X^i +\frac{\epsilon_{\alpha}^{\;\; \beta}}{\sqrt{h}}\left( B_{0i} \partial_{\beta} X^i + \partial_{\beta} \hat{X}^0 \right) \right].$$ Integrating the action over $V$ we get the dual action that has the same form as the original action except that the metric and the two-form field are now 
\begin{gather}
\tilde{g}_{00} = \frac{1}{g_{00}}, \notag \\ 
\tilde{g}_{0i} = \frac{B_{0i}}{g_{00}}, \qquad  \tilde{B}_{0i} = \frac{g_{0i}}{g_{00}},  \notag \\
\tilde{g}_{ij} = g_{ij}-\frac{g_{0i}g_{0j}-B_{0i}B_{0j}}{g_{00}}, \notag \\
\tilde{B}_{ij} = B_{ij}-\frac{g_{0i}B_{0j}-B_{0i}g_{0j}}{g_{00}}.
\label{eqn:tdual}
\end{gather} 
We would like to impose the condition that the T-dual theory is also conformally invariant. This can be imposed at 
one-loop using either the results of reference \cite{stringback}, i.e.\ equations \eqref{betaeqns}, or by considering the change in the measure of the path-integral. Either method suggests that the dilaton is shifted \cite{buscher1,*buscher2} to
\begin{equation}
\tilde{\phi}=\phi-\frac{1}{2} \log{g_{00}}.
\end{equation}

Note that if we take the special case of toroidal compactification on a flat background, then we get the well-known result that the radius of the compactification circle is inverted and the string coupling constant, which is the exponential of the expectation value of the dilaton, is modified by a factor of $\sqrt{\alpha'}/R.$ 

The argument given here is valid only for spherical worldsheets, hence the duality has only been proved to first order in string perturbation theory. By gauging the isometry the duality can be extended to higher genus worldsheets, but in this case the isometry orbits must be compact, or in other words the shift symmetry has to be along a compact coordinate \cite{rocek}.

Recently, Berkovits and Maldacena \cite{fermdual} have generalised B\"{u}scher's formulation of T-duality to the case where the worldsheet action is invariant under constant shifts of spacetime fermionic coordinates $\theta^{J},$ $J=1,\dots n.$ They show that under this duality the metric and the NS-NS 2-form potential do not change, and they give the transformation of the Ramond-Ramond fields in terms of the bispinor field strength.

In type IIA string theory the bispinor field strength is $$ F= \frac{1}{2} F^{(2)}_{\;\;\;\; a_1 a_2} \gamma^{a_1 a_2} + \frac{1}{4!}  F^{(4)}_{\;\;\;\; a_1 \dots a_4} \gamma^{a_1 \dots a_4} \gamma_{11},$$ where the 2-form $F^{(2)}$ and 4-form $F^{(4)}$ are the RR field strengths. In our notation, $\gamma-$matrices are 32 by 32 matrix representations of the ten-dimensional Clifford algebra; two Majorana-Weyl spinors describing $\mathcal{N}=2A$ supersymmetry are combined into a single Majorana-Dirac spinor. In type IIB string theory the Ramond-Ramond field strengths are the 1-form $F^{(1)},$ 3-form $F^{(3)}$ and the self-dual 5-form $F^{(5)},$ and we define the R-R bispinor field strength by $$ F= F^{(1)}_{\;\;\;\;a} \gamma^a \sigma^1 + \frac{1}{3!}  F^{(3)}_{\;\;\;\;a_1\dots a_3} \gamma^{a_1 \dots a_3} \left(i \sigma^2 \right) + \frac{1}{2.5!} F^{(5)}_{\;\;\;\;a_1\dots a_5} \gamma^{a_1 \dots a_5} \sigma^1.$$ In type IIB theory, two Majorana-Weyl spinors, $\varepsilon$ and $\hat{\varepsilon},$ with the same chirality are combined into an $SO(2)$ vector
$$\epsilon=\begin{pmatrix}
\varepsilon \\
\hat{\varepsilon}
\end{pmatrix},$$
which are rotated amongst each other by acting with $\e^{i \sigma^{2} \theta},$ and on which Pauli matrices act on in the obvious way.

Fermionic T-duality transforms the bispinor field strength in the following way:
\begin{equation}
\e^{\phi'} F'= \e^{\phi} F \pm 32 \sum_{I,J=1}^{N} \left(\varepsilon_{I} \otimes \hat{\varepsilon}_{J}\right) M_{IJ},
\label{eqn:BMFtrans}
\end{equation}
where we take $+$ in type IIB theory and $-$ in type IIA string theory. We will always combine two type IIA Weyl spinors into a Dirac spinor and type IIB spinors into an $SO(2)$ vector. However, in equation \eqref{eqn:BMFtrans}, and only there, the spinors are Weyl Killing spinors. 

Furthermore, under fermionic T-duality, the dilaton is transformed to  
\begin{equation}
\phi'=\phi + \frac{1}{2} \sum_{I=1}^{n}\left(\log \left(\frac{1}{2} M^{-1}\right) \right)_{II},
\label{eqn:BMdiltrans}
\end{equation}
where $M^{-1}$ satisfies $$ \partial_{a}  \left(M^{-1}\right)_{IJ} = 2 \bar{\epsilon}_{I} \gamma_{a} \gamma_{11} \epsilon_{J}$$ for type IIA theory, and $$ \partial_{a}  \left(M^{-1}\right)_{IJ} = 2 \bar{\epsilon}_{I} \gamma_{a} \sigma^{3} \epsilon_{J}$$ for type IIB theory---of course $\epsilon$ has a different meaning for each theory, as described above. The spinors, $\epsilon_{J},$ are Killing spinors corresponding to the constant shift symmetry of the fermionic coordinate $\theta^{J}$ in superspace $$ \theta_{I} \rightarrow \theta_{I} + \rho_{I},$$ where $\rho_{I}$ is a Grassmann-valued constant. Since this symmetry is abelian, $\{\epsilon_{I} Q_{I},\epsilon_{J} Q_{J}\}=0,$ where there is no summation over $I$ and $J.$ However, from the supersymmetry algebra $$\{\epsilon_{I} Q_{I},\epsilon_{J} Q_{J}\} = \bar{\epsilon}_{I} \gamma^{a} \epsilon_{J} P_{a},$$
 where $P$ is the generator for translations. Therefore, 
\begin{equation}
\bar{\epsilon}_{I} \gamma_{a} \epsilon_{J}=0,
\label{eqn:spincon}
\end{equation}
for all $I,J=1,\dots, n\in \mathbb{Z}^{+}.$ This condition can only be satisfied non-trivially for complex $\epsilon,$ so, in general, the transformation sends real supergravity backgrounds into complex backgrounds.

Fermionic T-duality preserves the number of supersymmetries, which is necessary in order for it to be a duality of string theory. Explicitly, the Killing spinors in the T-dual theory are 
\begin{equation}
\epsilon'_{I} = M_{IJ} \epsilon_{J}.
\label{newkillingspinors}
\end{equation}

\section{Type IIA supergravity symmetry} 
\label{symmIIA}

In our conventions, summarised in appendix \ref{conv}, the type IIA supergravity action is 
\begin{equation}
S= \frac{1}{2 \kappa^2} \int \d^{10} x \sqrt{g} \left\{ \e^{-2 \phi} \left[ R + 4 (\partial \phi)^2 - \frac{1}{12} H^2 \right] - \frac{1}{2} \left[ \frac{1}{2} F^{{(2)}^2} + \frac{1}{4!} F^{{(4)}^2 }\right] -\frac{1}{144} \frac{1}{\sqrt{g}} \epsilon \, \partial C^{(3)} \partial C^{(3)} B \right\}.
\end{equation}
The first square bracket is the action for the NSNS fields, the metric $g,$ the 2-form field $B,$ and the dilaton. The second set of terms constitute the action for the RR fields, the 1-form potential $C^{(1)},$ and the 3-form potential $C^{(3)}.$ The last term is the Chern-Simons term. The field strengths, $H, F^{(2)},F^{(4)}$ are defined by 
\begin{align*}
H &= \d B, \\
F^{(2)} &= \d C^{(1)}, \\
F^{(4)} &= \d C^{(3)} - H \wedge C^{(1)}. 
\end{align*}

The Bianchi identities for the field strengths are 
\begin{align}
&\d H =0, \label{bianchiH} \\
&\d F^{(2)}=0, \label{bianchi2}\\
&\d F^{(4)}-H \wedge F^{(2)}=0.
\label{bianchi4}
\end{align}
The equations of motion are 
\begin{align}
& \d \left(\e^{-2 \phi} \star H \right) + F^{(2)} \wedge \star F^{(4)} - \frac{1}{2} F^{(4)} \wedge F^{(4)} =0, \label{eomHA} \\
& \d  \star F^{(2)} + H \wedge \star F^{(4)} =0, \\
& \d  \star F^{(4)} - H \wedge F^{(4)} =0.
\end{align}
The Einstein equation is 
\begin{multline}
R_{ab}= - \frac{1}{4} g_{ab} \Box \phi + \frac{1}{2} g_{ab} \left(\partial \phi \right)^{2} - 2 \nabla_{a} \nabla_{b} \phi + \frac{1}{4} \left( H_{acd}H_{b}^{\;cd} - \frac{1}{12} g_{ab} H^{2} \right)\\
+\frac{1}{2} \e^{2 \phi} \left( F^{(2)}_{\;\;\;\;ac} F^{(2)\;c}_{\;\;\;\;b} -\frac{1}{16} g_{ab} F^{(2)^{2}} \right)  +\frac{1}{12} \e^{2 \phi} \left(F^{(4)}_{\;\;\;\;acde} F^{(4)\;cde}_{\;\;\;\;b} -\frac{3}{32} g_{ab} F^{(4)^{2}} \right),
\label{einstein}
\end{multline}
and the dilaton equation of motion is
\begin{equation}
R + 4 \Box \phi - 4 \left(\partial \phi \right)^{2} - \frac{1}{12} H^{2}=0.
\label{dileom}
\end{equation}
We can check the consistency of these equations by showing that the contracted Bianchi identity holds. Indeed, using the Bianchi identities and equations of motion for the field strengths and the Einstein equation, 
$$ \nabla^{a}(R_{ab}-\frac{1}{2} g_{ab}R) = - \partial_{b} \phi \left(R + 4 \Box \phi - 4 \left(\partial \phi \right)^{2} - \frac{1}{12} H^{2} \right),$$
which vanishes by the dilaton equation of motion. 

The Killing spinor equations from the variations of the gravitino and dilatino are 
\begin{equation} 
\nabla_{a} \epsilon - \frac{1}{8} H_{abc} \gamma^{bc} \gamma_{11} \epsilon - \frac{1}{16} \e^{\phi} F^{(2)}_{\;\;\;\;bc}\gamma^{bc} \gamma_{a} \gamma_{11} \epsilon + \frac{1}{192} \e^{\phi} F^{(4)}_{\;\;\;\;bcde} \gamma^{bcde}\gamma_{a} \epsilon=0,
\label{eqn:ksegrav}
\end{equation}
\begin{equation}
\left( \gamma^{a} \partial_{a} \phi - \frac{1}{12} H_{abc} \gamma^{abc} \gamma_{11} - \frac{3}{8} \e^{\phi} F^{(2)}_{\;\;\;\;ab} \gamma^{ab} \gamma_{11} + \frac{1}{96} \e^{\phi} F^{(4)} _{\;\;\;\; abcd} \gamma^{abcd} \right) \epsilon=0,
\label{eqn:ksedil}
\end{equation}
respectively. 

We will consider a transformation in the RR fields only. We also allow the dilaton to transform because at the quantum level this restores the conformal invariance of the string sigma model. For a string theory to be quantum mechanically consistent, it can be shown that the background fields must satisfy the supergravity equations of motion by imposing the vanishing of the beta-function \cite{stringback, betaGS}, or imposing $\kappa-$invariance of the Green-Schwarz action \cite{kappa, kappainvII}. The other NSNS fields, the metric and the 3-form field strength $H,$ are invariant under the transformation.

We consider the most general ansatz for the transformation of the fields:
\begin{align}
\e^{\phi} F^{(2)}_{\;\;\;\;ab} &\rightarrow \e^{\phi'} F'^{(2)}_{\;\;\;\;ab} = \e^{\phi} F^{(2)}_{\;\;\;\;ab} + \bar{\epsilon}_{I} \gamma_{ab} (S_1 + S_2 \gamma_{11} ) \eta_{J} M_{IJ}, \notag \\
\e^{\phi} F^{(4)}_{\;\;\;\;abcd} &\rightarrow \e^{\phi'} F'^{(4)}_{\;\;\;\;abcd} = \e^{\phi} F^{(4)}_{\;\;\;\;abcd} + \bar{\epsilon}_{I} \gamma_{abcd} (S_3 + S_4 \gamma_{11} ) \eta_{J} M_{IJ}, \label{fieldtrans}
\end{align}
where $\phi', M_{IJ}, S_{1} , \dots, S_{4}$ are arbitrary functions, and spinors $\epsilon_{I}, \eta_I,$ $I=1, \dots, n,$  satisfy the gravitino and dilatino Killing spinor equations. Both of the RR field strengths transform with the same spinors and functions, for if they transformed with different spinors and functions, then, for example, upon requiring that the Bianchi identity for the transformed 4-form field strength, equation \eqref{bianchi4}, holds, they would be identified. We will identify them from the onset in the interests of clarity and terseness. Furthermore, from equation \eqref{eomHA}, $\phi'$ can be identified as the transformed dilaton, which we will write as $$ \phi' = \phi +X,$$ where $X$ is an unknown function.

Let us consider each Bianchi identity and equation of motion in turn. First, consider the transformed Bianchi identity for the RR 2-form, equation \eqref{bianchi2}. Using the gravitino Killing spinor equation
\begin{align}
\nabla_{[a} F'^{(2)}_{\;\;\;\;bc]} &=  \nabla_{[a} \left( \e^{-X} F^{(2)}_{\;\;\;\;bc]}+\e^{-(\phi+X)} \bar{\epsilon}_{I}  \gamma_{bc]} \left(S_1 + S_2 \gamma_{11} \right) \eta_{J} M_{IJ} \right) \notag \\
&= -\e^{-X}  \left( F^{(2)}_{\;\;\;\;[bc}+\e^{-\phi} \bar{\epsilon}_{I}  \gamma_{[bc} \left(S_1 + S_2 \gamma_{11} \right) \eta_{J} M_{IJ} \right) \partial_{a]} X\notag \\
&  \qquad -  \e^{-(\phi+X)}  \bar{\epsilon}_{I} \gamma_{[bc} \left(S_1 + S_2 \gamma_{11} \right) \eta_{J} M_{IJ}\partial_{a]} \phi \notag \\
& \qquad \quad + \e^{-X} \epsilon_{I}^{\alpha} \eta_{J}^{\beta} \left( \frac{1}{4} \e^{-\phi} H_{de[a} \left( S_1 \left(\gamma_{bc]} \gamma^{de} \gamma_{11} \right)_{(\alpha \beta)} +S_2 \left( \gamma_{bc]} \gamma^{de} \right)_{[\alpha \beta]} \right) \right. \notag \\ 
& \qquad \qquad  + \frac{1}{8} F^{(2)}_{\;\;\;\;de} \left( S_1 \left(\gamma_{[bc} \gamma^{de} \gamma_{a]} \gamma_{11} \right)_{(\alpha \beta)} - S_2 \left( \gamma_{[bc} \gamma^{de} \gamma_{a]}  \right)_{[\alpha \beta]} \right) \notag\\
& \qquad \qquad \quad \biggl. - \frac{1}{96} F^{(4)}_{\;\;\;\;defg} \left( S_1 \left(\gamma_{[bc} \gamma^{defg} \gamma_{a]} \gamma_{11} \right)_{(\alpha \beta)} - S_2 \left( \gamma_{[bc} \gamma^{defg} \gamma_{a]}  \right)_{[\alpha \beta]} \right) \biggr) M_{IJ} \notag \\
& \qquad \qquad \qquad  + \e^{-(\phi+X)} \bar{\epsilon}_{I}  \gamma_{[bc} \partial_{a]} \left[ \left(S_1 + S_2 \gamma_{11} \right) M_{IJ} \right] \eta_{J}. 
\label{eqn:bian2}
\end{align}
The Greek indices, $\alpha, \beta =1, \dots, 32,$ are spinor indices. 

Now we can use the dilatino Killing spinor equation to express the term involving the derivative of the dilaton in the expression above in terms of the field strengths. As $\epsilon_{I}$ and $\eta_{J}$ satisfy the dilatino Killing spinor equation, \eqref{eqn:ksedil},
\begin{equation}
\epsilon_{I}^{\alpha} \gamma_{abc} \left( \gamma^{d} \partial_{d} \phi - \frac{1}{12} H_{def} \gamma^{def} \gamma_{11} - \frac{3 }{8} \e^{\phi} F^{(2)}_{\;\;\;\;de} \gamma^{de} \gamma_{11} + \frac{1}{96} \e^{\phi} F^{(4)}_{\;\;\;\;defg} \gamma^{defg} \right)^{(\alpha \beta)} \eta_{J}^{\beta}=0.
\label{eqn:dileqnb2}
\end{equation}
Using \footnote{This identity can be proved by using induction on $m$ with $n=1$ and then by induction on $n.$} \begin{equation}
\gamma_{a_1 \dots a_m} \gamma^{b_1 \dots b_n} = \sum_{k=0}^{\textup{min}(m,n)} c_{mn}^k \gamma_{[a_1 \dots a_{m-k}}^{\qquad \quad \;\; [b_1 \dots b_{n-k}} \delta_{a_{m-k+1}}^{b_{n-k+1}} \dots \delta_{a_m]}^{b_n]},
\label{gammaidentity}
\end{equation}
where $$ c_{mn}^k=(-1)^{kn+\frac{1}{2} k(k+1)} \frac{m! n!}{k!(m-k)!(n-k)!},$$
and \begin{align}
\left( \gamma^{a_1 \dots a_n} \right)^{ \alpha \beta} &= (-1)^{\frac{1}{2}n(n+1)+1} \left( \gamma^{a_1 \dots a_n} \right)^{ \beta \alpha }, \notag \\
\left( \gamma^{a_1 \dots a_n} \gamma_{11} \right)^{ \alpha \beta} &= (-1)^{\frac{1}{2}n(n+3)} \left( \gamma^{a_1 \dots a_n} \gamma_{11} \right)^{ \beta \alpha },
\label{eqn:symgamma}
\end{align}
equation \eqref{eqn:dileqnb2} implies that 
\begin{equation*}
\begin{split}
& \bar{\epsilon}_{I}  \gamma_{[bc} \eta_{J} \partial_{a]} \phi - \frac{1}{12} H_{de[a} \bar{\epsilon}_{I}  \left( 3 \gamma_{bc]}^{\;\;\;\;de} - 2 \delta_{b}^{d} \delta_{c]}^{e} \right) \gamma_{11} \eta_{J}  \\
& \qquad \qquad  - \frac{1}{8} \e^{\phi} F^{(2)}_{\;\;\;\;de} \bar{\epsilon}_{I} \left( \gamma_{abc}^{\;\;\;\;\;de} - 6 \gamma_{[a} \delta_{b}^{d} \delta_{c]}^{e} \right) \gamma_{11} \eta_{J} + \frac{1}{24} \e^{\phi} F^{(4)}_{\;\;\;\;defg} \bar{\epsilon}_{I} \left(\gamma_{[ab}^{\;\;\;\;def} \delta_{c]}^{g} -2 \gamma^{d} \delta_{a}^{e} \delta_{b}^{f} \delta_{c}^{g} \right) \eta_{J}=0. 
\end{split}
\end{equation*}
Similarly,
\begin{equation*}
\begin{split}
& \bar{\epsilon}_{I}  \gamma_{[bc} \gamma_{11} \eta_{J} \partial_{a]} \phi - \frac{1}{12} H_{de[a} \bar{\epsilon}_{I}  \left( 3 \gamma_{bc]}^{\;\;\;\;de} - 2 \delta_{b}^{d} \delta_{c]}^{e} \right) \eta_{J} - \frac{3}{4} \e^{\phi} F^{(2)}_{\;\;\;\;de} \bar{\epsilon}_{I}  \gamma_{[bc}^{\;\;\;\;d} \delta_{a]}^{e} \eta_{J} \\
& \qquad \qquad \qquad \qquad \qquad \qquad \qquad \qquad \qquad - \frac{1}{288} \e^{\phi} F^{(4)}_{\;\;\;\;defg} \bar{\epsilon}_{I} \left(\gamma_{abc}^{\;\;\;\;\;defg} -36 \gamma_{[b}^{\;\;d e} \delta_{c}^{f} \delta_{a]}^{g} \right) \gamma_{11}  \eta_{J}=0.
\end{split}
\end{equation*}

Substituting the two equations above in equation \eqref{eqn:bian2} and using the gamma matrix identities \eqref{gammaidentity} and \eqref{eqn:symgamma}, equation \eqref{eqn:bian2} becomes
\begin{align}
&\nabla_{[a} F'^{(2)}_{\;\;\;\;bc]} \notag \\
= &-\e^{-X} \left( F^{(2)}_{\;\;\;\;[bc}+\e^{-\phi} \bar{\epsilon}_{I}  \gamma_{[bc} \left(S_1 + S_2 \gamma_{11} \right) \eta_{J} M_{IJ} \right) \partial_{a]} X + \e^{-(\phi+X)} \bar{\epsilon}_{I}  \gamma_{[bc} \partial_{a]} \left[ \left(S_1 + S_2 \gamma_{11} \right) M_{IJ} \right] \eta_{J} \notag \\
& \quad - \frac{1}{3} \e^{-(\phi+X)} H_{abc} \bar{\epsilon}_{I} \left( S_1  \gamma_{11}+S_2  \right) \eta_{J} M_{IJ} + \frac{1}{2} \e^{-X} F^{(2)}_{\;\;\;\;de} \bar{\epsilon}_{I} \left( 2 S_1 \gamma_{[a} \delta_{b}^{d} \delta_{c]}^{e} \gamma_{11} - S_2 \gamma_{[bc}^{\;\;\;\;d} \delta_{a]}^{e} \right) \eta_{J} M_{IJ}\notag \\ 
& \qquad \quad + \frac{1}{144} \e^{-X} F^{(4)}_{\;\;\;\;defg} \bar{\epsilon}_{I} \left(48 S_1 \gamma^{d} \delta_{a}^{e} \delta_{b}^{f} \delta_{c}^{g} + S_2 \left( \gamma_{abc}^{\;\;\;\;\;defg} + 36 \gamma_{[a}^{\;\;d e} \delta_{b}^{f} \delta_{c]}^{g} \right) \right) \eta_{J} M_{IJ}. \notag
\label{eqn:bian22}
\end{align}
The expression above must vanish for the transformed Bianchi identity to be satisfied. Since we are considering generic supergravity solutions, by looking at the terms proportional to the RR 2-form field strength we conclude that \begin{equation} \partial_{a} X= S_1 \bar{\epsilon}_{I} \gamma_{a} \gamma_{11} \eta_{J} M_{IJ} \quad \textup{and} \quad S_2 \bar{\epsilon}_{I} \gamma_{abc} \eta_{J} M_{IJ} =0. \label{eqn:bian2con2} \end{equation}
From the terms proportional to the NSNS field strength we get that \begin{equation} \bar{\epsilon}_{I} \left( S_1 \gamma_{11} +S_2 \right) \eta_{J} M_{IJ} =0, \label{eqn:bian2conH} \end{equation} and from the terms involving the RR 4-form field strength we get \begin{equation} S_1 \bar{\epsilon}_{I} \gamma_{a} \eta_{J} M_{IJ}=0 \quad \textup{and} \quad S_2 \bar{\epsilon}_{I} \gamma_{abc} \gamma_{11} \eta_{J} M_{IJ} =0, \label{eqn:bian2con4} \end{equation} using $S_2 \bar{\epsilon}_{I} \gamma_{abc} \eta_{J} M_{IJ} =0,$ from equation \eqref{eqn:bian2con2}. Finally from the remaining terms we have that \begin{equation} \bar{\epsilon}_{I}  \gamma_{[bc}\Bigl( \partial_{a]} X \left(S_1 + S_2 \gamma_{11} \right) M_{IJ} - \partial_{a]} \bigl[ \left(S_1 + S_2 \gamma_{11} \right) M_{IJ} \bigr] \Bigr) \eta_{J}=0. \label{eqn:bian2condm} \end{equation}

Using similar techniques to those used above, we can also show that 
\begin{align}
\nabla_{[a} F'^{(4)}_{\;\;\;\;bcde]} -2 H_{[abc} F'^{(2)}_{\;\;\;\;de]} &= -\e^{-X}  \left( F^{(4)}_{\;\;\;\;[bcde}+\e^{-\phi} \bar{\epsilon}_{I}  \gamma_{[bcde} \left(S_3 + S_4 \gamma_{11} \right) \eta_{J} M_{IJ} \right) \partial_{a]} X\notag \\
& \;\; - 2 \e^{-(\phi+X)} H_{[abc} \bar{\epsilon}_{I} \gamma_{de]} \left( \left( S_2+ S_3 \right) \gamma_{11} + \left( S_1 + S_4 \right) \right) \eta_{J} M_{IJ} \notag \\ 
& \quad + \frac{1}{20} \e^{-X} F^{(2)}_{\;\;\;\;fg} \bar{\epsilon}_{I} S_3 \left(\gamma_{abcde}^{\quad \;\;\; fg}  - 20 \gamma_{[abc} \delta_{d}^{f} \delta_{e]}^{g} \right) \gamma_{11} \eta_{J} M_{IJ}  \notag\\
& \;\;\;\quad + \frac{1}{120} \e^{-X} F^{(4)}_{\;\;\;\;fghi} \bar{\epsilon}_{I} \left(10 S_3 \left( \gamma_{[bcde}^{\quad \;\;fgh} \delta_{a]}^{i} + 12 \gamma_{[ab}^{\;\;\;\;f} \delta_{c}^{g} \delta_{d}^{h} \delta_{e]}^{i} \right) \right. \notag \\
& \qquad \qquad \qquad \qquad \qquad \left. + S_4 \left( \gamma_{abcde}^{\quad \;\;\; fghi} - 120 \gamma_{[a} \delta_{b}^{f} \delta_{c}^{g} \delta_{d}^{h} \delta_{e]}^{i} \right) \gamma_{11} \right) \eta_{J} M_{IJ} \notag \\
& \qquad \qquad + \e^{-(\phi+X)} \bar{\epsilon}_{I}  \gamma_{[bcde} \partial_{a]} \left[ \left(S_3 + S_4 \gamma_{11} \right) M_{IJ} \right] \eta_{J},
\label{eqn:bian4}
\end{align}
\begin{align}
& \nabla^{b} F'^{(2)}_{\;\;\;\;ba} - \frac{1}{6} H^{bcd} F'^{(4)}_{\;\;\;\;abcd} \notag \\ 
= &-\e^{-X}  \left( F^{(2)}_{\;\;\;\;ba}+\e^{-\phi} \bar{\epsilon}_{I}  \gamma_{ba} \left(S_1 + S_2 \gamma_{11} \right) \eta_{J} M_{IJ} \right) \partial^{b} X + \e^{-(\phi+X)} \bar{\epsilon}_{I}  \gamma_{ba} \partial^{b} \left[ \left(S_1 + S_2 \gamma_{11} \right) M_{IJ} \right] \eta_{J}\notag \\
& \qquad \quad - \frac{1}{6} \e^{-(\phi+X)} H_{bcd}  \bar{\epsilon}_{I}\gamma_{a}^{\;\;bcd} \left( (S_1 + S_4) \gamma_{11}+ (S_2 + S_3) \right) \eta_{J} M_{IJ} \notag \\ 
& \qquad\qquad \quad  + \frac{1}{4} \e^{-X} F^{(2)}_{\;\;\;\;bc} \bar{\epsilon}_{I} \left( 4 S_1 \gamma^{b} \delta_{a}^{c} \gamma_{11} + S_2 \gamma_{a}^{\;\;bc} \right) \eta_{J} M_{IJ} - \frac{1}{6} \e^{-X} F^{(4)}_{\;\;\;\;bcda} \bar{\epsilon}_{I} S_2 \gamma^{bcd} \eta_{J} M_{IJ} 
\label{eom2}
\end{align}
and
\begin{align}
& \nabla^{d} F'^{(4)}_{\;\;\;\;dabc} - \frac{1}{144} \epsilon_{a b c d_1 \dots d_7} H^{d_1 d_2 d_3} F'^{(4) d_4 \dots d_7} \notag \\ 
= &-\e^{-X} \left( F^{(4)}_{\;\;\;\;dabc}+\e^{-\phi} \bar{\epsilon}_{I}  \gamma_{dabc} \left(S_3 + S_4 \gamma_{11} \right) \eta_{J} M_{IJ} \right) \partial^{d} X \notag \\
& \quad + \e^{-(\phi+X)} \bar{\epsilon}_{I}  \gamma_{dabc} \partial^{d} \left[ \left(S_3 + S_4 \gamma_{11} \right) M_{IJ} \right] \eta_{J} + \frac{3}{2} \e^{-X} F^{(2)}_{\;\;\;\;d[c} \bar{\epsilon}_{I} \left( S_3 \gamma_{ab]}^{\quad d}  \gamma_{11} - 2 S_4 \gamma_{a} \delta_{b]}^{d} \right)  \eta_{J} M_{IJ}  \notag\\
& \qquad - \frac{1}{48} \e^{-X} F^{(4)}_{\;\;\;\;defg} \bar{\epsilon}_{I} \left( S_3 \left( \gamma_{abc}^{\quad \;defg} + 36 \gamma_{[a}^{\;\;\;de} \delta_{b}^{f} \delta_{c]}^{g} \right) + 48 S_4 \gamma^{d} \delta_{a}^{e} \delta_{b}^{f} \delta_{c}^{g} \gamma_{11} \right) \eta_{J} M_{IJ}. 
\label{eom4}
\end{align}
For the transformed equation of motion for the RR 4-form field to hold, from equation \eqref{eom4}, the following must be satisfied \begin{gather}
\partial_{a} X= -S_4 \bar{\epsilon}_{I} \gamma_{a} \gamma_{11} \eta_{J} M_{IJ}, \label{eom4condX} \\
S_4 \bar{\epsilon}_{I} \gamma_{a} \eta_{J} M_{IJ} =0, \label{eom4conds41gamma} \\
S_3 \bar{\epsilon}_{I} \gamma_{abc} \eta_{J} M_{IJ} =0, \qquad S_3 \bar{\epsilon}_{I} \gamma_{abc} \gamma_{11} \eta_{J} M_{IJ} =0, \label{eom4conds3gamma} \\
\bar{\epsilon}_{I}  \gamma_{abcd}\Bigl( \partial^{d} X \left(S_3 + S_4 \gamma_{11}  \right) M_{IJ} - \partial^{d} \bigl[ \left(S_3 + S_4 \gamma_{11} \right) M_{IJ} \bigr] \Bigr) \eta_{J}=0. \label{eom4con}
\end{gather}
Note that if, for example, $$ \bar{\epsilon}_{I} \gamma_{abc} \gamma_{11} \eta_{J} M_{IJ} =0, $$ then $$\bar{\epsilon}_{I} \gamma_{a_1 \dots a_7} \eta_{J} M_{IJ} =0,$$ for
\begin{equation}
\gamma_{a_1 \dots a_m}= -\frac{(-1)^{\frac{1}{2}(10-m)(10-m+1)}}{(10-m)!} \epsilon_{a_1 \dots a_m b_1 \dots b_{10-m}}  \gamma^{b_1 \dots b_{10-m}} \gamma_{11},
\label{eqn:gammadual}
\end{equation}
proved in Appendix \ref{gammadual}.

Now, if equations \eqref{eqn:bian2con2}, (\ref{eom4condX}--\ref{eom4conds3gamma}) hold, then the expressions in equations \eqref{eqn:bian4} and \eqref{eom2} vanish only if
\begin{gather}
(S_1 +S_4) \bar{\epsilon}_{I} \gamma_{ab} \eta_{J} M_{IJ}+(S_2 +S_3) \bar{\epsilon}_{I} \gamma_{ab} \gamma_{11} \eta_{J} M_{IJ}=0, \\
\bar{\epsilon}_{I}  \gamma_{[bcde}\Bigl( \partial_{a]} X \left(S_3 + S_4 \gamma_{11} \right) M_{IJ} - \partial_{a]} \bigl[ \left(S_3 + S_4 \gamma_{11} \right) M_{IJ} \bigr] \Bigr) \eta_{J}=0
\label{eqn:bian4con2}
\end{gather}
and
\begin{gather}
(S_1 +S_4) \bar{\epsilon}_{I} \gamma_{abcd} \gamma_{11} \eta_{J} M_{IJ}+(S_2 +S_3) \bar{\epsilon}_{I} \gamma_{abcd} \eta_{J} M_{IJ}=0, \\
\bar{\epsilon}_{I}  \gamma_{ab}\Bigl( \partial^{b} X \left(S_1 + S_2 \gamma_{11} \right) M_{IJ} - \partial^{b} \bigl[ \left(S_1 + S_2 \gamma_{11} \right) M_{IJ} \bigr] \Bigr) \eta_{J}=0, \label{eom2con}
\end{gather}
respectively.

In summary, the Killing spinors and functions that describe the transformation must satisfy 
\begin{gather}
\partial_{a} X= S_1 \bar{\epsilon}_{I} \gamma_{a} \gamma_{11} \eta_{J} M_{IJ}, \qquad  (S_1 +S_4) \bar{\epsilon}_{I} \gamma_{a} \gamma_{11} \eta_{J} M_{IJ}=0, \notag \\
\bar{\epsilon}_{I} \left( S_1 \gamma_{11} +S_2 \right) \eta_{J} M_{IJ} =0, \notag \\
S_{1} \bar{\epsilon}_{I} \gamma_{a} \eta_{J} M_{IJ}=0, \qquad S_{4} \bar{\epsilon}_{I} \gamma_{a} \eta_{J} M_{IJ}=0, \notag \\
S_2 \bar{\epsilon}_{I} \gamma_{abc} \eta_{J} M_{IJ} =0, \qquad S_2 \bar{\epsilon}_{I} \gamma_{abc} \gamma_{11} \eta_{J} M_{IJ} =0, \notag \\
S_3 \bar{\epsilon}_{I} \gamma_{abc} \eta_{J} M_{IJ} =0, \qquad S_3 \bar{\epsilon}_{I} \gamma_{abc} \gamma_{11} \eta_{J} M_{IJ} =0, \notag \\
(S_1 +S_4) \bar{\epsilon}_{I} \gamma_{ab} \eta_{J} M_{IJ}+ (S_2 +S_3) \bar{\epsilon}_{I} \gamma_{ab} \gamma_{11} \eta_{J} M_{IJ}=0, \notag \\
(S_1 +S_4) \bar{\epsilon}_{I} \gamma_{abcd} \gamma_{11} \eta_{J} M_{IJ}+ (S_2 +S_3) \bar{\epsilon}_{I} \gamma_{abcd} \eta_{J} M_{IJ}=0, \notag \\
\bar{\epsilon}_{I}  \gamma_{[bc}\Bigl( \partial_{a]} X \left(S_1 + S_2 \gamma_{11} \right) M_{IJ} - \partial_{a]} \bigl[ \left(S_1 + S_2 \gamma_{11} \right) M_{IJ} \bigr] \Bigr) \eta_{J}=0,\notag \\
\bar{\epsilon}_{I}  \gamma_{ab}\Bigl( \partial^{b} X \left(S_1 + S_2 \gamma_{11} \right) M_{IJ} - \partial^{b} \bigl[ \left(S_1 + S_2 \gamma_{11} \right) M_{IJ} \bigr] \Bigr) \eta_{J}=0, \notag \\
\bar{\epsilon}_{I}  \gamma_{[bcde}\Bigl( \partial_{a]} X \left(S_3 + S_4 \gamma_{11}   \right) M_{IJ} - \partial_{a]} \bigl[ \left(S_3 + S_4 \gamma_{11}   \right) M_{IJ} \bigr] \Bigr) \eta_{J}=0 \notag \\
\bar{\epsilon}_{I}  \gamma_{abcd}\Bigl( \partial^{d} X \left(S_3 + S_4 \gamma_{11}   \right) M_{IJ} - \partial^{d} \bigl[ \left(S_3 + S_4 \gamma_{11}   \right) M_{IJ} \bigr] \Bigr) \eta_{J}=0 
\label{RRcon}
\end{gather}
in order for the Bianchi identities and equations of motion for the transformed RR fields to be satisfied.

Let us consider the NSNS 3-form equations. The Bianchi identity for the NSNS 3-form field is invariant under the transformation, so we do not need to consider it. However, using the Killing spinor equations and equations \eqref{RRcon}, the equation of motion for the NSNS 3-form, \eqref{eomHA}, reduces to 
\begin{align}
&  \e^{-(\phi + 2 X)} F_{cd} \bar{\epsilon}_{I} \left( \left(3 S_{3} -S_{4} \gamma_{11} \right)\delta_{a}^{c} \delta_{b}^{d} - S_{3} \gamma_{ab}^{\;\;\;cd}\right) \eta_{J} M_{IJ} \notag \\
& \qquad -  \e^{-(\phi + 2 X)}S_{3}  F_{abcd} \bar{\epsilon}_{I} \gamma^{cd} \gamma_{11} \eta_{J} M_{IJ} - 4 \e^{-2(\phi + X)} \bar{\epsilon}_{I} \gamma_{[a} \eta_{J} \partial_{b]} \left(S_4 M_{IJ}\right) \notag \\
& \qquad\qquad  + \frac{1}{2} \e^{-2(\phi + X)} \left( \bar{\epsilon}_{I} \gamma_{abcd} \left(S_3  + S_4 \gamma_{11} \right) \eta_{J} \right) \left( \bar{\epsilon}_{K} \gamma^{cd} \left(S_1  + S_2 \gamma_{11}\right) \eta_{L} \right) M_{IJ} M_{KL} \notag \\
& \qquad \qquad \qquad \;\; - \frac{1}{48} \e^{-2(\phi + X)} \left( \bar{\epsilon}_{I} \gamma_{cdef} \left(S_3  + S_4 \gamma_{11} \right) \eta_{J} \right) ( \bar{\epsilon}_{K} \gamma_{ab}^{\;\;\;cdef} \left(S_3 \gamma_{11} + S_4  \right) \eta_{L} ) M_{IJ} M_{KL}=0.
\label{eqn:transHeqnA}
\end{align}
The supergravity fields we are considering are generic, so, in particular from the term proportional to the RR 2-form field, we must have that 
$$S_{3} \bar{\epsilon}_{I} \gamma_{abcd} \eta_{J} M_{IJ}=0,$$
which implies that $S_{3} =0,$ for this is precisely the combination that enters in the transformation of the 4-form RR field strength. Furthermore, since $S_3=0,$ from $$(S_1 +S_4) \bar{\epsilon}_{I} \gamma_{ab} \eta_{J} M_{IJ}+ (S_2 +S_3) \bar{\epsilon}_{I} \gamma_{ab} \gamma_{11} \eta_{J} M_{IJ}=0$$ we get that $$ S_2 \bar{\epsilon}_{I} \gamma_{ab} \gamma_{11} \eta_{J} M_{IJ} \propto  \bar{\epsilon}_{I} \gamma_{ab} \eta_{J} M_{IJ},$$ hence without loss of generality we can set $S_2=0.$ 

Since $S_2$ and $S_3$ vanish, we must have that at least one of  $$\bar{\epsilon}_{I} \gamma_{ab} \eta_{J} M_{IJ}, \qquad \bar{\epsilon}_{I} \gamma_{abcd} \gamma_{11} \eta_{J} M_{IJ}$$ are non-zero in order for the transformation to be non-trivial. Therefore, using equations 
\begin{gather*}
(S_1 +S_4) \bar{\epsilon}_{I} \gamma_{ab} \eta_{J} M_{IJ}=0,\qquad (S_1 +S_4) \bar{\epsilon}_{I} \gamma_{abcd} \gamma_{11} \eta_{J} M_{IJ}=0,
\end{gather*} 
from the set of equations \eqref{RRcon} with $S_2=S_3=0,$ we deduce that $$S_4=-S_1.$$ Without loss of generality we can let $S_1=1.$

Furthermore, from the first term in equation \eqref{eqn:transHeqnA}, the spinors must satisfy
\begin{equation}
\bar{\epsilon}_{I}\gamma_{11} \eta_{J} M_{IJ}=0.
\label{eqn:gamma11con}
\end{equation}

The last two terms in equation \eqref{eqn:transHeqnA} are quartic in spinors and they can be simplified using Fierz identities.

The Fierz identity for commuting spinors $\lambda, \chi, \psi, \varphi$ in $d-$dimensions is
\[
\left( \bar{\lambda} M \chi \right) \left( \bar{\psi} N \varphi \right) = 2^{-[d/2]} \sum_{I} \left( \bar{\lambda} M \mathcal{O}^I N \varphi \right) \left( \bar{\psi} \mathcal{O}_I \chi \right),
\]
where $M,N$ are arbitrary combination of gamma matrices and $$\{\mathcal{O}_I\}= \{ \mathbb{I}, \gamma_a, i \gamma_{ab}, i \gamma_{abc}, \gamma_{abcd}, \dots\}$$ forms a basis for $2^{[d/2]} \times 2^{[d/2]} $ matrices and $$\{\mathcal{O}^I\}= \{ \mathbb{I}, \gamma^a, i \gamma^{ab}, i \gamma^{abc}, \gamma^{abcd}, \dots\}$$ is the dual basis.

Using Fierz identities, equation \eqref{eqn:transHeqnA} with 
\begin{gather*}
S_1=-S_4=1, \quad S_2=S_3=0, \\
\bar{\epsilon}_{I}\gamma_{11} \eta_{J} M_{IJ}=0
\end{gather*}
becomes
\begin{align*}
& 4 \bar{\epsilon}_{I} \gamma_{[a} \eta_{J} \partial_{b]} M_{IJ} - \biggl( 16  \left( \bar{\epsilon}_{I} \gamma_{[a} \gamma_{11} \eta_{J} \right) \left( \bar{\epsilon}_{K} \gamma_{b]}  \eta_{L} \right) -  \left( \bar{\epsilon}_{I} \gamma_{ab} \eta_{J} \right) \left( \bar{\epsilon}_{K} \gamma_{11} \eta_{L} \right) + \left( \bar{\epsilon}_{I} \gamma_{ab} \gamma_{11} \eta_{J} \right) \left( \bar{\epsilon}_{K} \eta_{L} \right)\biggr. \notag \\
& \qquad \qquad\quad +\frac{1}{2} \left( \bar{\epsilon}_{I} \gamma_{abcd}  \eta_{J} \right) \left( \bar{\epsilon}_{K} \gamma^{cd} \gamma_{11} \eta_{L} \right)+\frac{1}{48} \left( \bar{\epsilon}_{I} \gamma_{abcdef} \gamma_{11} \eta_{J} \right) \left( \bar{\epsilon}_{K} \gamma^{cdef}  \eta_{L} \right) \biggr) M_{IJ} M_{KL}=0.
\end{align*}
We can use equation \eqref{eqn:gamma11con} again to simplify the above equation to 
\begin{align}
& 4 \bar{\epsilon}_{I} \gamma_{[a} \eta_{J} \partial_{b]} M_{IJ} - \biggl( 16  \left( \bar{\epsilon}_{I} \gamma_{[a} \gamma_{11} \eta_{J} \right) \left( \bar{\epsilon}_{K} \gamma_{b]}  \eta_{L} \right) + \left( \bar{\epsilon}_{I} \gamma_{ab} \gamma_{11} \eta_{J} \right) \left( \bar{\epsilon}_{K} \eta_{L} \right)\biggr. \notag \\
& \qquad \qquad \qquad \left. +\frac{1}{2} \left( \bar{\epsilon}_{I} \gamma_{abcd}  \eta_{J} \right) \left( \bar{\epsilon}_{K} \gamma^{cd} \gamma_{11} \eta_{L} \right) +\frac{1}{48} \left( \bar{\epsilon}_{I} \gamma_{abcdef} \gamma_{11} \eta_{J} \right) \left( \bar{\epsilon}_{K} \gamma^{cdef}  \eta_{L} \right) \right) M_{IJ} M_{KL}=0.
\label{Hcon}
\end{align}

So far, having only the dilaton and Einstein equation to consider, we have the following conditions on the Killing spinors and functions in the transformation of the fields:
\begin{gather}
S_1=-S_4=1, \label{eqn:S11con} \\ 
S_3=S_2=0, \label{eqn:S20con} \\ 
\partial_{a} X=  \bar{\epsilon}_{I} \gamma_{a} \gamma_{11} \eta_{J} M_{IJ}, \label{eqn:dXcon} \\
\bar{\epsilon}_{I} \gamma_{11} \eta_{J} M_{IJ} =0, \label{eqn:gamma11con2} \\
\bar{\epsilon}_{I} \gamma_{a} \eta_{J} M_{IJ}=0,  \label{eqn:gamma10con} \\
\bar{\epsilon}_{I}  \gamma_{[bc} \eta_{J} \left( \partial_{a]} X  M_{IJ} - \partial_{a]} M_{IJ} \right) =0, \label{eqn:quarb2con} \\
\bar{\epsilon}_{I}  \gamma_{ab} \eta_{J} \left( \partial^{b} X  M_{IJ} - \partial^{b} M_{IJ} \right)=0, \label{eqn:quare2con} \\
\bar{\epsilon}_{I}  \gamma_{[bcde} \gamma_{11} \eta_{J} \Bigl( \partial_{a]} X  M_{IJ} - \partial_{a]} M_{IJ} \Bigr) =0, \label{eqn:quarb4con} \\
\bar{\epsilon}_{I}  \gamma_{abcd} \gamma_{11} \eta_{J} \left( \partial^{d} X  M_{IJ} - \partial^{d} M_{IJ} \right)=0 \label{eqn:quare4con}
\end{gather}
and equation \eqref{Hcon}.

The Dilaton equation for the transformed fields is 
\begin{align*}
R + 4 \Box \phi' - 4 \left(\partial \phi' \right)^{2} - \frac{1}{12} H^{2}=0,
\end{align*}
which using the dilaton equation for the original fields implies that 
\begin{equation}
\Box X= 2 \partial_{a} \phi \partial^{a} X + \partial_{a} X \partial^{a} X.
\label{dilcon}
\end{equation}
Using equation \eqref{eqn:dXcon} and the Killing spinor equation from the variation of the gravitino,
\begin{align*}
\Box X &= \nabla_{a} \left(  \bar{\epsilon}_{I} \gamma^{a} \gamma_{11} \eta_{J} M_{IJ} \right) \\
&= - \frac{3}{4} \e^{\phi} F_{bc}  \bar{\epsilon}_{I} \gamma^{bc} \eta_{J} M_{IJ} + \frac{1}{48} \e^{\phi} F_{bcde}  \bar{\epsilon}_{I} \gamma^{bcde} \gamma_{11} \eta_{J} M_{IJ} +  \bar{\epsilon}_{I} \gamma^{a} \gamma_{11} \eta_{J} \partial_{a} M_{IJ}.
\end{align*}
However, since $\epsilon_{I}$ and $\eta_{I},$ for all $I= 1 \dots, n,$  satisfy the dilatino Killing spinor equation, \eqref{eqn:ksedil}, 
\begin{equation*}
\begin{split} &\epsilon_{I}^{\alpha} \left( \gamma_{11} \left( \gamma^{a} \partial_{a} \phi - \frac{1}{12} H_{abc} \gamma^{abc} \gamma_{11} - \frac{3}{8} \e^{\phi} F^{(2)}_{\;\;\;\;ab} \gamma^{ab} \gamma_{11}  + \frac{1}{96} \e^{\phi} F^{(4)} _{\;\;\;\; abcd} \gamma^{abcd} \right) \right)_{(\alpha \beta)} \eta_{J}^{\beta} =0,
\end{split}
\end{equation*}
hence, using equation \eqref{eqn:symgamma},
$$- \frac{3}{4} \e^{\phi} F_{bc}  \bar{\epsilon}_{I} \gamma^{bc} \eta_{J} M_{IJ} + \frac{1}{48} \e^{\phi} F_{bcde}  \bar{\epsilon}_{I} \gamma^{bcde} \gamma_{11} \eta_{J} M_{IJ} = 2 \bar{\epsilon}_{I} \gamma^{a} \gamma_{11} \eta_{J} \partial_{a} \phi M_{IJ}.$$ 
Therefore,
\begin{align*}
\Box X &= 2 \bar{\epsilon}_{I} \gamma^{a} \gamma_{11} \eta_{J} \partial_{a} \phi M_{IJ} +  \bar{\epsilon}_{I} \gamma^{a} \gamma_{11} \eta_{J} \partial_{a} M_{IJ}.
\end{align*}
So, from equation \eqref{dilcon}, the transformed dilaton equation is satisfied if 
\begin{equation}
\bar{\epsilon}_{I} \gamma^{a} \gamma_{11} \eta_{J} \left( \bar{\epsilon}_{K} \gamma_{a} \gamma_{11} \eta_{L} M_{IJ} M_{KL} - \partial_{a} M_{IJ} \right) =0.
\label{dilconquar}
\end{equation}

Finally, we have to find conditions on the spinors and functions in the transformation in order for the Einstein equation to be satisfied for the transformed fields. Using equations \eqref{eqn:S11con}, \eqref{eqn:S20con} and \eqref{dilcon} and the Einstein equation for the original fields, Einstein's equation becomes
\begin{align}
&\frac{1}{4} g_{ab} \Box X -2 \nabla_{a} \nabla_{b} X + \e^{\phi} \left( F^{(2)\;\;\,c}_{\;\;\;\;(a} \bar{\epsilon}_{I} \gamma_{b)c}  \eta_{J} -\frac{1}{16} g_{ab} F^{(2)}_{\;\;\;\;cd} \bar{\epsilon}_{I} \gamma^{cd} \eta_{J} \right) M_{IJ} - \frac{1}{6} \e^{\phi} \left( F^{(4)\;\;\,cde}_{\;\;\;\;(a} \bar{\epsilon}_{I} \gamma_{b)cde} \gamma_{11} \eta_{J} \right. \notag \\ 
& \;  \left.- \frac{3}{32} g_{ab} F^{(4)}_{\;\;\;\;cdef} \bar{\epsilon}_{I} \gamma^{cdef} \gamma_{11} \eta_{J}  \right) M_{IJ} + \frac{1}{2} \left( \left(\bar{\epsilon}_{I} \gamma_{ac} \eta_{J} \right) \left( \bar{\epsilon}_{K} \gamma_{b}^{\;\;c} \eta_{L} \right)  -\frac{1}{16} g_{ab} \left(\bar{\epsilon}_{I} \gamma_{cd} \eta_{J} \right) \left(\bar{\epsilon}_{K} \gamma^{cd} \eta_{L} \right) \right) M_{IJ} M_{KL}\notag \\
& \qquad + \frac{1}{12} \bigg( \left(\bar{\epsilon}_{I} \gamma_{acde} \gamma_{11} \eta_{J} \right) \left( \bar{\epsilon}_{K} \gamma_{b}^{\;\;cde} \gamma_{11} \eta_{L} \right)  -\frac{3}{32} g_{ab} \left(\bar{\epsilon}_{I} \gamma_{cdef} \gamma_{11} \eta_{J} \right) \left(\bar{\epsilon}_{K} \gamma^{cdef} \gamma_{11} \eta_{L} \right) \bigg) M_{IJ} M_{KL} =0. 
\label{eqn:einst}
\end{align} 
Now, consider 
\begin{equation*}
\begin{split} &\bar{\epsilon}_{I} \left[ \gamma_{(a|} \gamma_{11} \left( \nabla_{|b)} - \frac{1}{8} H_{|b)cd} \gamma^{cd} \gamma_{11} - \frac{1}{16} \e^{\phi} F^{(2)}_{\;\;\;\;cd} \gamma^{cd} \gamma_{|b)} \gamma_{11} + \frac{1}{192} \e^{\phi} F^{(4)} _{\;\;\;\;cdef} \gamma^{cdef} \gamma_{|b)} \right) \right] \eta_{J} =0.
\end{split}
\end{equation*}
Adding this to the same expression, but with $\epsilon_{I}$ and $\eta_{J}$ interchanged we get 
\begin{equation*}
\begin{split}
&  \bar{\epsilon}_{I}  \gamma_{(a} \gamma_{11} \nabla_{b)} \eta_{J} + \bar{\eta}_{J}  \gamma_{(a} \gamma_{11} \nabla_{b)} \epsilon_{I} + \frac{1}{8} \e^{\phi} F^{(2)}_{\;\;\;\;cd} \bar{\epsilon}_{I} \left( 4 \gamma_{(a}^{\;\;c} \delta_{b)}^{d} +  g_{ab} \gamma^{cd} \right) \eta_{J}  \\ 
& \qquad \qquad\qquad\qquad \qquad\qquad\quad\qquad \qquad \qquad- \frac{1}{96} \e^{\phi} F^{(4)}_{\;\;\;\;cdef} \bar{\epsilon}_{I} \left( 8 \gamma_{(a}^{\;\;cde} \delta_{b)}^{f} +  g_{ab} \gamma^{cdef} \right) \gamma_{11} \eta_{J} =0,
\end{split}
\end{equation*}
using equations \eqref{gammaidentity} and \eqref{eqn:symgamma}. The above equation and the equation obtained by contracting the two free indices in the above equation can be used to reduce equation \eqref{eqn:einst} to 
\begin{align}
&\frac{1}{4} g_{ab} \Box X -2 \nabla_{a} \nabla_{b} X + 2 \nabla_{(a} \left( \bar{\epsilon}_{I}  \gamma_{b)} \gamma_{11}  \eta_{J} \right) M_{IJ} - \frac{1}{4} g_{ab} \nabla_{c} \left( \bar{\epsilon}_{I}  \gamma^{c} \gamma_{11}  \eta_{J} \right) M_{IJ}  \notag \\ 
& \quad + \frac{1}{2} \left( \left(\bar{\epsilon}_{I} \gamma_{ac} \eta_{J} \right) \left( \bar{\epsilon}_{K} \gamma_{b}^{\;\;c} \eta_{L} \right)  -\frac{1}{16} g_{ab} \left(\bar{\epsilon}_{I} \gamma_{cd} \eta_{J} \right) \left(\bar{\epsilon}_{K} \gamma^{cd} \eta_{L} \right) \right) M_{IJ} M_{KL} \notag \\
& \qquad  + \frac{1}{12} \bigg( \left(\bar{\epsilon}_{I} \gamma_{acde} \gamma_{11} \eta_{J} \right) \left( \bar{\epsilon}_{K} \gamma_{b}^{\;\;cde} \gamma_{11} \eta_{L} \right) -\frac{3}{32} g_{ab} \left(\bar{\epsilon}_{I} \gamma_{cdef} \gamma_{11} \eta_{J} \right) \left(\bar{\epsilon}_{K} \gamma^{cdef} \gamma_{11} \eta_{L} \right) \biggr) M_{IJ} M_{KL} =0. 
\label{eqn:einst2}
\end{align} 
Using equation \eqref{eqn:dXcon} to simplify the terms on the first line, Fierz identities and equations (\ref{eqn:gamma11con2}) and (\ref{eqn:gamma10con}), it can be shown that equation \eqref{eqn:einst2} is 
\begin{align}
& - 2 \bar{\epsilon}_{I} \gamma_{(a} \gamma_{11} \eta_{J} \left( \partial_{b)} M_{IJ} + 2 \bar{\epsilon}_{K} \gamma_{b)} \gamma_{11} \eta_{L} M_{IL} M_{KJ} \right) + \frac{1}{4} g_{ab} \bar{\epsilon}_{I} \gamma^{c} \gamma_{11} \eta_{J} \left( \partial_{c} M_{IJ} + 2 \bar{\epsilon}_{K} \gamma_{c} \gamma_{11} \eta_{L} M_{IL} M_{KJ} \right)\notag\\
& \qquad +\biggl( 4 \left( \bar{\epsilon}_{I} \gamma_{a} \eta_{L} \right) \left(\bar{\epsilon}_{K} \gamma_{b} \eta_{J} \right) + \frac{1}{2} \left( \bar{\epsilon}_{I} \gamma_{a}^{\;\;c} \gamma_{11} \eta_{J} \right) \left(\bar{\epsilon}_{K} \gamma_{bc} \gamma_{11} \eta_{L} \right) + \frac{1}{12} \left( \bar{\epsilon}_{I} \gamma_{a}^{\;\;cde} \eta_{J} \right) \left(\bar{\epsilon}_{K} \gamma_{bcde} \eta_{L} \right)\notag \\
&  \qquad \qquad  + \frac{1}{16}  g_{ab} \left( \bar{\epsilon}_{I} \eta_{J} \right) \left(\bar{\epsilon}_{K} \eta_{L} \right) - \frac{1}{2} g_{ab} \left( \bar{\epsilon}_{I} \gamma_{c} \eta_{L} \right) \left(\bar{\epsilon}_{K} \gamma^{c} \eta_{J} \right) - \frac{1}{32} g_{ab} \left( \bar{\epsilon}_{I} \gamma_{cd} \gamma_{11} \eta_{J} \right) \left(\bar{\epsilon}_{K} \gamma^{cd} \gamma_{11} \eta_{L} \right)\notag\\
% & - \frac{1}{16} g_{ab} \left( \bar{\epsilon}_{I} \gamma_{11} \eta_{J} \right) \left(\bar{\epsilon}_{K} \gamma_{11} \eta_{L} \right) \notag\\
& \qquad \qquad \qquad \qquad \qquad  \qquad \qquad \Bigl. -\frac{1}{128} g_{ab} \left( \bar{\epsilon}_{I} \gamma_{cdef} \eta_{J} \right) \left(\bar{\epsilon}_{K} \gamma^{cdef} \eta_{L} \right) \biggr)  M_{IJ} M_{KL}=0.
\label{eqn:quarEcon}
\end{align}
This is a quartic condition on the spinors. Moreover, from equations \eqref{Hcon}, (\ref{eqn:quarb2con}--\ref{eqn:quare4con}) and equation \eqref{dilconquar} we also have that the spinors must satisfy
\begin{align}
& 4\bar{\epsilon}_{I} \gamma_{[a} \eta_{J} \left( \partial_{b]} M_{IJ} - 4 \bar{\epsilon}_{K} \gamma_{b]} \gamma_{11} \eta_{L} M_{IL} M_{KJ} \right) - \biggl( \left( \bar{\epsilon}_{I} \gamma_{ab} \gamma_{11} \eta_{J} \right) \left( \bar{\epsilon}_{K} \eta_{L} \right) \biggr. \notag \\
& \qquad \quad\qquad \left. +\frac{1}{2} \left( \bar{\epsilon}_{I} \gamma_{abcd}  \eta_{J} \right) \left( \bar{\epsilon}_{K} \gamma^{cd} \gamma_{11} \eta_{L} \right) +\frac{1}{48} \left( \bar{\epsilon}_{I} \gamma_{abcdef} \gamma_{11} \eta_{J} \right) \left( \bar{\epsilon}_{K} \gamma^{cdef}  \eta_{L} \right) \right) M_{IJ} M_{KL}=0, 
\label{Hcon2} 
\end{align}
\begin{align}
& \bar{\epsilon}_{I} \gamma_{[ab} \eta_{J} \left( \partial_{c]} M_{IJ} + 2 \bar{\epsilon}_{K} \gamma_{c]} \gamma_{11} \eta_{L} M_{IL} M_{KJ} \right) + \biggl( \frac{1}{2} \left( \bar{\epsilon}_{I}  \gamma_{[ab}^{\;\;\;\;d} \eta_{J} \right) \left( \bar{\epsilon}_{K} \gamma_{c]d} \gamma_{11} \eta_{L} \right)\notag \\
& \qquad \quad  + \frac{1}{6} \left( \bar{\epsilon}_{I}  \gamma_{abc} \gamma_{11} \eta_{J} \right) \left( \bar{\epsilon}_{K} \eta_{L} \right) - \frac{1}{4} \left( \bar{\epsilon}_{I}  \gamma_{[ab}^{\;\;\;\;de} \eta_{J} \right) \left( \bar{\epsilon}_{K} \gamma_{c]de} \gamma_{11} \eta_{L} \right) + \frac{1}{3} \left( \bar{\epsilon}_{I}  \gamma_{abcd} \gamma_{11} \eta_{J} \right) \left( \bar{\epsilon}_{K} \gamma^{d} \eta_{L} \right)\notag\\
& \qquad \qquad \qquad +\frac{2}{3} \left( \bar{\epsilon}_{I}  \gamma_{abcd} \gamma_{11} \eta_{L} \right) \left( \bar{\epsilon}_{K}  \gamma^{d} \eta_{J} \right) \left. - \frac{1}{36} \left( \bar{\epsilon}_{I}  \gamma_{abcdef} \gamma_{11} \eta_{J} \right) \left( \bar{\epsilon}_{K} \gamma^{def} \eta_{L} \right) \right) M_{IJ} M_{KL}  =0, \label{eqn:quarb2con2} 
\end{align}
\begin{align}
& \bar{\epsilon}_{I} \gamma_{ab} \eta_{J} \left( \partial^{b} M_{IJ} + 2 \bar{\epsilon}_{K} \gamma^{b} \gamma_{11} \eta_{L} M_{IL} M_{KJ} \right) + \biggl( \left( \bar{\epsilon}_{I}  \gamma_{a} \eta_{J} \right) \left( \bar{\epsilon}_{K} \gamma_{11} \eta_{L} \right) + 2 \left( \bar{\epsilon}_{I}  \gamma_{a} \eta_{L} \right) \left( \bar{\epsilon}_{K} \gamma_{11}  \eta_{J} \right)\notag \\
& \qquad \qquad\qquad \qquad -\frac{3}{4} \left( \bar{\epsilon}_{I}  \gamma_{abc} \eta_{J} \right) \left( \bar{\epsilon}_{K} \gamma^{bc} \gamma_{11} \eta_{L} \right) \left. - \frac{1}{12} \left( \bar{\epsilon}_{I}  \gamma_{abcd} \eta_{J} \right) \left( \bar{\epsilon}_{K} \gamma^{bcd} \gamma_{11} \eta_{L} \right) \right) M_{IJ} M_{KL} =0, \label{eqn:quare2con2} 
\end{align}
\begin{align}
& \bar{\epsilon}_{I} \gamma_{[abcd} \gamma_{11} \eta_{J} \left( \partial_{e]} M_{IJ} + 2 \bar{\epsilon}_{K} \gamma_{e]} \gamma_{11} \eta_{L} M_{IL} M_{KJ} \right)+ \biggl( \left( \bar{\epsilon}_{I}  \gamma_{[abc} \gamma_{11} \eta_{J} \right) \left( \bar{\epsilon}_{K} \gamma_{de]} \gamma_{11} \eta_{L} \right) \notag \\
& \quad  + \left( \bar{\epsilon}_{I}  \gamma_{[abc}^{\;\;\;\;\;\;f} \eta_{J} \right) \left( \bar{\epsilon}_{K} \gamma_{de]f} \eta_{L} \right)  + \frac{1}{5} \left( \bar{\epsilon}_{I}  \gamma_{abcdef} \eta_{J} \right) \left( \bar{\epsilon}_{K} \gamma^{f} \eta_{L} \right) + \frac{2}{5} \left( \bar{\epsilon}_{I}  \gamma_{abcdef} \eta_{L} \right) \left( \bar{\epsilon}_{K} \gamma^{f} \eta_{J} \right)\notag\\
& \qquad \quad\left. -\frac{1}{4} \left( \bar{\epsilon}_{I}  \gamma_{[abcd}^{\quad \;\;\; fg} \gamma_{11} \eta_{J} \right) \left( \bar{\epsilon}_{K}  \gamma_{e]fg} \gamma_{11} \eta_{L} \right)  + \frac{1}{20} \left( \bar{\epsilon}_{I}  \gamma_{abcdefg} \gamma_{11} \eta_{J} \right) \left( \bar{\epsilon}_{K} \gamma^{fg} \gamma_{11} \eta_{L} \right) \right) M_{IJ} M_{KL} =0, 
\label{eqn:quarb4con2} 
\end{align}
\begin{align}
& \bar{\epsilon}_{I} \gamma_{abcd} \gamma_{11} \eta_{J} \left( \partial^{d} M_{IJ} + 2 \bar{\epsilon}_{K} \gamma^{d} \gamma_{11} \eta_{L} M_{IL} M_{KJ} \right) + \biggl(3  \left( \bar{\epsilon}_{I}  \gamma_{[ab} \eta_{J} \right) \left( \bar{\epsilon}_{K} \gamma_{c]} \eta_{L} \right) \notag \\
& \qquad + 6 \left( \bar{\epsilon}_{I}  \gamma_{[ab} \eta_{L} \right) \left( \bar{\epsilon}_{K} \gamma_{c]}  \eta_{J} \right) + \frac{1}{2} \left( \bar{\epsilon}_{I}  \gamma_{abc} \eta_{J} \right) \left( \bar{\epsilon}_{K} \eta_{L} \right) + \frac{3}{2} \left( \bar{\epsilon}_{I}  \gamma_{[ab}^{\;\;\;\;\;d} \gamma_{11} \eta_{J} \right) \left( \bar{\epsilon}_{K} \gamma_{c]d} \gamma_{11} \eta_{L} \right)  \notag\\
& \qquad \qquad \left. - \frac{3}{4} \left( \bar{\epsilon}_{I}  \gamma_{[ab}^{\;\;\;\;de} \eta_{J} \right) \left( \bar{\epsilon}_{K} \gamma_{c]de} \eta_{L} \right) - \frac{1}{12} \left( \bar{\epsilon}_{I}  \gamma_{abcdef} \gamma_{11} \eta_{J} \right) \left( \bar{\epsilon}_{K} \gamma^{def} \gamma_{11} \eta_{L} \right) \right) M_{IJ} M_{KL} =0, 
\label{eqn:quare4con2} 
\end{align}
\begin{align}
& \bar{\epsilon}_{I} \gamma^{a} \gamma_{11} \eta_{J} \left( \partial_{a} M_{IJ} + 2 \bar{\epsilon}_{K} \gamma_{a} \gamma_{11} \eta_{L} M_{IL} M_{KJ} \right) + \biggl( \left( \bar{\epsilon}_{I}  \gamma^{a} \eta_{J} \right) \left( \bar{\epsilon}_{K} \gamma_{a} \eta_{L} \right) + 2 \left( \bar{\epsilon}_{I}  \gamma^{a} \eta_{L} \right) \left( \bar{\epsilon}_{K} \gamma_{a}  \eta_{J} \right) \notag \\
& \qquad \qquad\qquad \left. -\frac{1}{12} \left( \bar{\epsilon}_{I}  \gamma^{abc} \eta_{J} \right) \left( \bar{\epsilon}_{K} \gamma_{abc} \eta_{L} \right) - \frac{1}{12} \left( \bar{\epsilon}_{I}  \gamma^{abc} \gamma_{11} \eta_{J} \right) \left( \bar{\epsilon}_{K} \gamma_{abc} \gamma_{11} \eta_{L} \right) \right) M_{IJ} M_{KL} =0, 
\label{dilconquar2}
\end{align}
respectively, where Fierz identities have been used to rewrite equations (\ref{eqn:quarb2con}--\ref{eqn:quare4con}) and equation \eqref{dilconquar}.

The only set of quadratic constraints on the spinors that we have found that solve equations (\ref{eqn:quarEcon}--\ref{dilconquar2}) is
\begin{gather*}
\bar{\epsilon}_{I} \gamma_{a} \eta_{J}=0, \quad \bar{\epsilon}_{I} \eta_{J} M_{IJ} =0,  \quad \bar{\epsilon}_{I} \gamma_{ab} \gamma_{11} \eta_{J} M_{IJ}=0 \\
\bar{\epsilon}_{I} \gamma_{abc} \eta_{J} M_{IJ}=0, \quad  \bar{\epsilon}_{I} \gamma_{abc} \gamma_{11} \eta_{J} M_{IJ}=0, \quad \bar{\epsilon}_{I} \gamma_{cdef} \eta_{J} M_{IJ}=0,  \\
\partial_{a} M_{IJ} = - 2 \bar{\epsilon}_{K} \gamma_{a} \gamma_{11} \eta_{L} M_{IL} M_{KJ}.
\end{gather*}

We have shown that the type IIA supergravity equations admit a symmetry described by the following transformations of the dilaton and RR field strengths
\begin{align}
\phi \rightarrow \phi' &= \phi + X,\notag \\
\e^{\phi} F^{(2)}_{\;\;\;\;ab} \rightarrow \e^{\phi'} F'^{(2)}_{\;\;\;\;ab} &= \e^{\phi} F^{(2)}_{\;\;\;\;ab} + \bar{\epsilon}_{I} \gamma_{ab} \eta_{J} M_{IJ}, \notag \\
\e^{\phi} F^{(4)}_{\;\;\;\;abcd} \rightarrow \e^{\phi'} F'^{(4)}_{\;\;\;\;abcd} &= \e^{\phi} F^{(4)}_{\;\;\;\;abcd} - \bar{\epsilon}_{I} \gamma_{abcd} \gamma_{11} \eta_{J} M_{IJ},
\label{eqn:IIAsymm}
\end{align}
where the Killing spinors must satisfy
 \begin{gather}
\bar{\epsilon}_{I} \gamma_{a} \eta_{J}=0, \label{eqn2:gamma10con} \\
\bar{\epsilon}_{I} \gamma_{11} \eta_{J} M_{IJ} =0,  \\
\bar{\epsilon}_{I} \eta_{J} M_{IJ} =0, \quad \bar{\epsilon}_{I} \gamma_{ab} \gamma_{11} \eta_{J} M_{IJ}=0, \label{con2A1}\\  \bar{\epsilon}_{I} \gamma_{abc} \eta_{J} M_{IJ}=0, \quad  \bar{\epsilon}_{I} \gamma_{abc} \gamma_{11} \eta_{J} M_{IJ}=0, \quad \bar{\epsilon}_{I} \gamma_{cdef} \eta_{J} M_{IJ}=0, \label{con2A2}\\
\partial_{a} X=  \bar{\epsilon}_{I} \gamma_{a} \gamma_{11} \eta_{J} M_{IJ}, \label{eqn:Xdef} \\
\partial_{a} M_{IJ} = - 2 \bar{\epsilon}_{K} \gamma_{a} \gamma_{11} \eta_{L} M_{IL} M_{KJ}. \label{eqn:Mdef}
\end{gather} 
Equation \eqref{eqn:Mdef} is equivalent to 
\begin{equation}
\partial_{a} (M^{-1})_{IJ} =  2 \bar{\epsilon}_{J} \gamma_{a} \gamma_{11} \eta_{I},
\label{eqn:Minvdef}
\end{equation}
and equation \eqref{eqn:Xdef} can be solved to find $X$ up to a constant of integration:
\begin{equation}
X =  \frac{1}{2}  \sum_{I=1}^{n}\left(\log M^{-1} \right)_{II}.
\end{equation} 

The integrability conditions arising from equations \eqref{eqn:Xdef} and \eqref{eqn:Minvdef} are trivial because  
\begin{align*}
 \nabla_{[a} \nabla_{b]} X = \frac{1}{2} H_{abc} \bar{\epsilon}_{I} \gamma^{c} \eta_{J} M_{IJ} \quad \textup{and} \quad 
 \nabla_{[a} \nabla_{b]}  (M^{-1})_{IJ}= H_{abc} \bar{\epsilon}_{J} \gamma^{c} \eta_{I},
\end{align*}
which vanish by equation \eqref{eqn2:gamma10con}.

In the transformations given by Berkovits and Maldacena the spinors $\epsilon_{I}$ and $\eta_{I}$ are identified. This is sufficient for 
\begin{gather*}
\bar{\epsilon}_{I} \eta_{J} M_{IJ} =0, \quad \bar{\epsilon}_{I} \gamma_{ab} \gamma_{11} \eta_{J} M_{IJ}=0 ,\\  \bar{\epsilon}_{I} \gamma_{abc} \eta_{J} M_{IJ}=0, \quad  \bar{\epsilon}_{I} \gamma_{abc} \gamma_{11} \eta_{J} M_{IJ}=0, \quad \bar{\epsilon}_{I} \gamma_{cdef} \eta_{J} M_{IJ}=0.
\end{gather*} 
When $\epsilon_{I}$ and $\eta_{I}$ are identified, only the symmetric part of $M_{IJ}$ contributes in the transformations of the fields, so without loss of generality we can let $M_{IJ}$ be symmetric in $I$ and $J,$ as a consequence of which the above equations are satisfied. If we identity  $\epsilon_{I}$ and $\eta_{I}$ then we recover the transformations of Berkovits and Maldacena, but with an extra condition on the spinors, namely that $$ \bar{\epsilon}_{I} \gamma_{11} \epsilon_{J} M_{IJ} =0.$$ 

When $n=1,$ we can explicitly show that the solution to \begin{gather*} \bar{\epsilon} \eta =0, \quad \bar{\epsilon} \gamma_{ab} \gamma_{11} \eta=0 , \\ \quad \bar{\epsilon} \gamma_{abc} \eta=0, \quad  \bar{\epsilon} \gamma_{abc} \gamma_{11} \eta=0, \quad \bar{\epsilon} \gamma_{cdef} \eta=0 \end{gather*} is $$\epsilon \propto \eta.$$ However, when $n>1,$ these conditions do not reduce to the transformation rules of fermionic T-duality.

\section{Type IIB supergravity symmetry} 
\label{symmIIB}

The type IIB supergravity action is \begin{multline}
S= \frac{1}{2 \kappa^2} \int \d^{10} x \sqrt{g} \left\{ \e^{-2 \phi} \left[ R + 4 (\partial \phi)^2 - \frac{1}{12} H^2 \right] \right.\\ 
\left.- \frac{1}{2} \left[ F^{{(1)}^2} + \frac{1}{3!} F^{{(3)}^2} + \frac{1}{2.5!} F^{{(5)}^2} \right] - \frac{1}{192} \frac{1}{\sqrt{g}} \epsilon \,  C^{(4)} \partial  B \partial C^{(2)} \right\}.
\end{multline}
In type IIB supergravity the RR fields are the scalar $C^{(0)},$ the 2-form $C^{(2)}$ and the 4-form $C^{(4)}.$ 
In terms of potentials $B,C^{(0)},C^{(2)}$ and $C^{(4)},$ the field strengths are defined to be
\begin{align*}
H = \, & \d B, \quad  F^{(1)} = \d C^{(0)},  \quad F^{(3)} = \d C^{(2)} - H C^{(0)}, \\
&F^{(5)} = \d C^{(4)} - \frac{1}{2} C^{(2)} \wedge H +\frac{1}{2} B \wedge \d C^{(2)}.
\end{align*}
The 5-form field strength is constrained to be self-dual.

The Bianchi identities for the field strengths are 
\begin{align}
&\d H =0, \label{bianchiHB} \\
&\d F^{(1)}=0,
\label{bianchi1B} \\
&\d F^{(3)}- H \wedge F^{(1)}=0,
\label{bianchi3B} \\
&\d F^{(5)}- H \wedge F^{(3)}=0.
\label{bianchi5B}
\end{align}
The equations of motion are 
\begin{align}
& \d \left(\e^{-2 \phi} \star H \right) - F^{(1)} \wedge \star F^{(3)} - F^{(3)} \wedge F^{(5)} =0, \\
& \d  \star F^{(1)} + H \wedge \star F^{(3)} =0, \\
& \d  \star F^{(3)} + H \wedge F^{(5)} =0.
\end{align}
The equation of motion for the 5-form field strength, $F^{(5)},$ is equivalent to the Bianchi identity for the 5-form, equation \eqref{bianchi5B}, as it is self-dual.
Moreover, the Einstein equation is 
\begin{multline}
R_{ab}= -\frac{1}{4} g_{ab} \Box \phi + \frac{1}{2} g_{ab} \left(\partial \phi \right)^{2} - 2 \nabla_{a} \nabla_{b} \phi 
+ \frac{1}{4} \left( H_{acd}H_{b}^{\;cd} - \frac{1}{12} g_{ab} H^{2} \right)  \\
+ \frac{1}{2} \e^{2 \phi} F^{(1)}_{\;\;\;\;a} F^{(1)}_{\;\;\;\;b}  + \frac{1}{4} \e^{2 \phi} \left( F^{(3)}_{\;\;\;\;acd} F^{(3)\;cd}_{\;\;\;\;b} -\frac{1}{12} g_{ab} F^{(3)^{2}} \right) + \frac{1}{96} \e^{2 \phi} F^{(5)}_{\;\;\;\;acdef} F^{(5)\;cdef}_{\;\;\;\;b},
\label{einsteinB}
\end{multline}
noting that $F^{(5)^2}$ vanishes because the 5-form field is self-dual. Finally, the dilaton equation of motion is the same as the type IIA supergravity dilaton equation of motion, equation \eqref{dileom}. Also, the twice-contracted Bianchi identity is again satisfied using the equations of motion for the fields. 

The Killing spinor equations from the variation of the gravitino and dilatino are 
\begin{equation} 
\nabla_{a} \epsilon - \frac{1}{8} H_{abc} \gamma^{bc} \sigma^{3} \epsilon - \frac{1}{8} \e^{\phi} \biggl( F^{(1)}_{\;\;\;\;b}\gamma^{b} \gamma_{a} \left( i \sigma^{2} \right) \epsilon  + \frac{1}{3!} F^{(3)}_{\;\;\;\;bcd}\gamma^{bcd} \gamma_{a} \sigma^{1} \epsilon + \frac{1}{2.5!} F^{(5)}_{\;\;\;\;bcdef} \gamma^{bcdef} \gamma_{a} \left( i \sigma^{2} \right) \epsilon \biggr) = 0,
\label{eqn:ksegravB}
\end{equation}
\begin{equation}
\left( \gamma^{a} \partial_{a} \phi - \frac{1}{12} H_{abc} \gamma^{abc} \sigma^{3} + \e^{\phi} F^{(1)}_{\;\;\;\;a} \gamma^{a} \left( i \sigma^{2} \right) + \frac{1}{12} \e^{\phi} F^{(3)} _{\;\;\;\; abc} \gamma^{abc} \sigma^{1} \right) \epsilon=0,
\label{eqn:ksedilB}
\end{equation}
respectively. 

We now consider the most general transformation of the RR field strengths and we will also allow the dilaton to transform:
\begin{align*}
\e^{\phi} F^{(1)}_{\;\;\;\;a} &\rightarrow \e^{\phi'} F'^{(1)}_{\;\;\;\;a}= \e^{\phi} F^{(1)}_{\;\;\;\;a} + \bar{\epsilon}_{I} \gamma_{a} S^{(1)} \eta_J M_{IJ}, \\
\e^{\phi} F^{(3)}_{\;\;\;\;abc} &\rightarrow \e^{\phi'} F'^{(3)}_{\;\;\;\;abc}= \e^{\phi} F^{(3)}_{\;\;\;\;abc} + \bar{\epsilon}_{I} \gamma_{abc} S^{(2)} \eta_J M_{IJ}, \\
\e^{\phi} F^{(5)}_{\;\;\;\;abcde} &\rightarrow \e^{\phi'} F'^{(5)}_{\;\;\;\;abcde}= \e^{\phi} F^{(5)}_{\;\;\;\;abcde} + \bar{\epsilon}_{I} \gamma_{abcde} S^{(3)} \eta_J M_{IJ},
\end{align*}
where $M_{IJ}$ is an arbitrary function; the spinors $\epsilon_I, \eta _I$ satisfy the gravitino and dilatino Killing spinor equations; $$S^{(1,2,3)}= \sum_{\mu} S_{\mu}^{(1,2,3)} \sigma^{ \star \mu},$$ $\sigma^{\star \mu}=\left( \mathbb{I}, \sigma^{1}, i \sigma^{2},\sigma^{3} \right)^{\mu}.$ The field $\phi'$ is some arbitrary field, which is identified with the transformed dilaton upon considering the NSNS 3-form equation of motion with transformed fields. We let $$ \phi'= \phi + X,$$ where $X$ is some arbitrary function.

Note that the RR fields need not \emph{a priori} transform with the same spinors and coefficients. However, as in section \ref{symmIIA}, if we let them transform with different spinors and coefficients, then we will find from equation (\ref{bianchi3B}), for example, that the spinors and functions have to be identified. 

As in section \ref{symmIIA}, we let the NSNS fields $g$ and  $H$ be invariant under the transformation.

It is important that the 5-form field strength remains self-dual after the transformation. The Hodge dual of $\delta F^{(5)}$ is 
\begin{align*}
\star \left( \bar{\epsilon}_{I} \gamma_{a_1 \dots a_5} S^{(3)} \eta_J M_{IJ} \d x^{a_1} \wedge \dots \wedge \d x^{a_5} \right) &=  \bar{\epsilon}_{I} \left( \frac{1}{5!} \epsilon_{a_1 \dots a_5 b_1 \dots b_5 }\gamma^{b_1 \dots b_5} \right) S^{(3)} \eta_J M_{IJ} \d x^{a_1} \wedge \dots \wedge \d x^{a_5} \\
&=  \bar{\epsilon}_{I}  \left( \gamma_{a_1 \dots a_5} \gamma_{11} \right) S^{(3)} \eta_J M_{IJ} \d x^{a_1} \wedge \dots \wedge \d x^{a_5},
\end{align*}
by identity \eqref{eqn:gammadual}. Hence if we let $\gamma_{11} \eta_J=\eta_J$ then the transformed 5-form field strength is self-dual. Recall that in type IIB supergravity all the Killing spinors have the same chirality, hence $\gamma_{11} \epsilon_I=\epsilon_I.$

We will now find the constraints that the various functions and the Killing spinors must satisfy so that the transformed fields satisfy the Bianchi identities and the equations of motion. First, let us consider the Bianchi identities. Using the gravitino Killing spinor equation, the Bianchi identity for the transformed RR 1-form field strength is 
\begin{align*}
\nabla_{[a} F'^{(1)}_{\;\;\;\;b]} &=\nabla_{[a} \left( \e^{-X} F^{(1)}_{\;\;\;\;b]} + \e^{-(\phi+X)} \bar{\epsilon}_{I} \gamma_{b]} S^{(1)} \eta_J M_{IJ}\right) \\
&= - \e^{-X} \partial_{[a} X \left(F_{b]} + \e^{-\phi} \bar{\epsilon}_I \gamma_{b]} S^{(1)} \eta_{J} M_{IJ} \right) - \e^{-(\phi+X)} \partial_{[a} \phi \bar{\epsilon}_I \gamma_{b]} S^{(1)} \eta_J M_{IJ} \\ 
& \qquad + \e^{-X}M_{IJ} \epsilon_{I}^{\alpha} \left( \frac{1}{8} \e^{- \phi} H_{cd[a} \left( \left(\gamma_{b]}^{\;\;cd} \right)_{\alpha \beta} \left( S^{(1)} \sigma^{3}\right) + \left(\gamma_{b]}^{\;\;cd} \right)_{\beta \alpha} \left(\sigma^{3} S^{(1)} \right)\right) \right. \\ 
& \qquad \quad + \frac{1}{8} F^{(1)}_{\;\;c} \left( \left(\gamma_{[b} \gamma^{c} \gamma_{a]} \right)_{\alpha \beta} \left(i S^{(1)} \sigma^{2} \right) - \left(\gamma_{[b} \gamma^{c} \gamma_{a]} \right)_{\beta \alpha} \left(i \sigma^{2} S^{(1)} \right) \right) \\
& \qquad \quad  + \frac{1}{8.3!} F^{(3)}_{\;\;cde} \left( \left(\gamma_{[b} \gamma^{cde} \gamma_{a]} \right)_{\alpha \beta} \left( S^{(1)} \sigma^{1} \right) + \left(\gamma_{[b} \gamma^{cde} \gamma_{a]} \right)_{\beta \alpha} \left( \sigma^{1} S^{(1)} \right) \right) \\
& \qquad \quad \left. + \frac{1}{16.5!} F^{(5)}_{\;\;cdefg} \left( \left(\gamma_{[b} \gamma^{cdefg} \gamma_{a]} \right)_{\alpha \beta} \left(i S^{(1)} \sigma^{2} \right) - \left(\gamma_{[b} \gamma^{cdefg} \gamma_{a]} \right)_{\beta \alpha} \left(i \sigma^{2} S^{(1)} \right) \right) \right) \eta_{J}^{\beta}   \\
& \qquad \qquad + \e^{-(\phi+X)} \bar{\epsilon}_I \gamma_{[b} \partial_{a]} S^{(1)} \eta_{J} M_{IJ} +\e^{-(\phi+X)} \bar{\epsilon}_I \gamma_{[b} S^{(1)} \eta_{J} \partial_{a]} M_{IJ}=0.
\end{align*}

Now, from the dilatino Killing spinor equation 
$$ \bar{\epsilon}_{I} \gamma_{ba} S^{(1)} \left( \gamma^{c} \partial_{c} \phi - \frac{1}{12} H_{cde} \gamma^{cde} \sigma^{3} + \e^{\phi} F^{(1)}_{\;\;\;\;c} \gamma^{c} (i \sigma^2) + \frac{1}{12} \e^{\phi} F^{(3)}_{\;\;\;\;cde} \gamma^{cde} \sigma^1 \right) \eta_{J}=0.$$ 
Adding the above equation to 
$$ \bar{\eta}_{J} \gamma_{ba} S^{(1) \textup{t}} \left( \gamma^{c} \partial_{c} \phi - \frac{1}{12} H_{cde} \gamma^{cde} \sigma^{3} + \e^{\phi} F^{(1)}_{\;\;\;\;c} \gamma^{c} (i \sigma^2) + \frac{1}{12} \e^{\phi} F^{(3)}_{\;\;\;\;cde} \gamma^{cde} \sigma^1 \right) \epsilon_{I}=0,$$ 
where $S^{\textup{t}}$ is the transpose of $S,$ and using the first identity in the set of equations \eqref{eqn:symgamma} we can show that
\begin{align*}
\bar{\epsilon}_{I} \partial_{[a} \phi \gamma_{b]}  S^{(1)} \eta_{J} = \epsilon_{I}^{\alpha} &\left(  \frac{1}{48} H_{cde} \left( \left(\gamma_{ba}\gamma^{cde} \right)_{\alpha \beta} \left( S^{(1)} \sigma^{3}\right) + \left(\gamma_{ba}\gamma^{cde}  \right)_{\beta \alpha} \left(\sigma^{3} S^{(1)} \right) \right) \right. \\ 
& \qquad -  \frac{1}{4}\e^{\phi} F^{(1)}_{\;\;c} \left( \left(\gamma_{ba} \gamma^{c}\right)_{\alpha \beta} \left( S^{(1)} i\sigma^{2} \right) - \left(\gamma_{ba} \gamma^{c} \right)_{\beta \alpha} \left(i \sigma^{2} S^{(1)} \right) \right) \\
& \qquad \qquad \left. - \frac{1}{48} \e^{\phi} F^{(3)}_{\;\;cde} \left( \left(\gamma_{ba} \gamma^{cde} \right)_{\alpha \beta} \left( S^{(1)} \sigma^{1} \right) + \left(\gamma_{ba} \gamma^{cde}\right)_{\beta \alpha} \left( \sigma^{1} S^{(1)} \right) \right) \right) \eta_J.
\end{align*}

Therefore, using the above equation and performing some gamma matrix manipulations, using equations \eqref{gammaidentity} and \eqref{eqn:symgamma}, we can show that
\begin{align*}
\nabla_{[a} F'^{(1)}_{\;\;\;\;b]} & = \e^{-(\phi+X)} \bar{\epsilon}_I \gamma_{[b} \Bigl( \partial_{a]}  \left(S^{(1)} M_{IJ} \right) - \partial_{a]} X  S^{(1)} M_{IJ} \Bigr) \eta_{J}  - \e^{-X}  \partial_{[a} X F^{(1)}_{\;\;b]} \\
& \quad - \frac{1}{24} \e^{-(\phi+X)} H_{cde} \bar{\epsilon}_{I} \left(\gamma_{ba}^{\;\;\;cde} + \gamma^{c} \delta^{d}_{b} \delta^{e}_{a} \right)  \left( S^{(1)}_{0} \sigma^{3} + S^{(1)}_{3} \mathbb{I} \right) \eta_{J} M_{IJ} \\ 
& \qquad + \frac{1}{4} \e^{-X} F^{(1)}_{\;\;c} \bar{\epsilon}_{I} \left( \gamma_{ba}^{\;\;\;c} \left(S^{(1)}_{0} i \sigma^{2} - S^{(1)}_{2} \mathbb{I} \right)  - 4 \gamma_{[b} \delta^{c}_{a]} \left(S^{(1)}_{1} \sigma^{3} - S^{(1)}_{3}  \sigma^{1} \right) \right) \eta_{J} M_{IJ}\\
& \qquad\quad   + \frac{1}{4} \e^{-X}  F^{(3)}_{\;\;cd[a} \bar{\epsilon}_{I} \left( \gamma_{b]}^{\;\;cd}  \left(S^{(1)}_{2} \sigma^{3} + S^{(1)}_{3} i \sigma^{2} \right)- 2 \gamma^{c} \delta^{d}_{b]} \left(S^{(1)}_{0} \sigma^{1} + S^{(1)}_{1} \mathbb{I} \right) \right) \eta_{J} M_{IJ} \\
& \qquad\qquad - \frac{1}{24} \e^{-X} F^{(5)}_{\;\;bacde} \bar{\epsilon}_{I} \gamma^{cde} \left(S^{(1)}_{0} i \sigma^{2} - S^{(1)}_{2} \mathbb{I} \right) \eta_{J} M_{IJ}.
\end{align*}
This expression must vanish if the transformed RR 1-form field strength is to satisfy the Bianchi identity. We are considering generic supergravity fields, so the expression vanishes only if
\begin{gather}
\partial_{a} X = \bar{\epsilon}_{I} \gamma_{a} \left(S^{(1)}_{1} \sigma^{3} - S^{(1)}_{3}  \sigma^{1} \right) \eta_{J} M_{IJ}, \label{eqn:bian1condX} \\
\bar{\epsilon}_{I} \gamma_{abcde} \left( S^{(1)}_{0} \sigma^{3} + S^{(1)}_{3} \mathbb{I} \right) \eta_{J} M_{IJ}=0, \\
\bar{\epsilon}_{I} \gamma_{a} \left( S^{(1)}_{0} \sigma^{3} + S^{(1)}_{3} \mathbb{I} \right) \eta_{J} M_{IJ}=0, \\
\bar{\epsilon}_{I} \gamma_{abc}\left(S^{(1)}_{0} i \sigma^{2} - S^{(1)}_{2} \mathbb{I} \right) \eta_{J} M_{IJ}=0, \\
\bar{\epsilon}_{I} \gamma_{abc}  \left(S^{(1)}_{2} \sigma^{3} + S^{(1)}_{3} i \sigma^{2} \right) \eta_{J} M_{IJ}=0, \\
\bar{\epsilon}_{I} \gamma_{a} \left(S^{(1)}_{0} \sigma^{1} + S^{(1)}_{1} \mathbb{I} \right) \eta_{J} M_{IJ}=0, \\
\bar{\epsilon}_I \gamma_{[b} \Bigl( \partial_{a]}  \left(S^{(1)} M_{IJ} \right) - \partial_{a]} X  S^{(1)} M_{IJ} \Bigr) \eta_{J}=0.
\end{gather}

We can also show that 
\begin{align*}
&\nabla_{[a} F'^{(3)}_{\;\;\;\;bcd]} + F'^{(1)}_{\;\;\;\;[a} H_{bcd]} \\ 
= &\e^{-(\phi+X)} \bar{\epsilon}_I \gamma_{[bcd} \Bigl( \partial_{a]}  \left(S^{(2)} M_{IJ} \right) - \partial_{a]} X  S^{(2)} M_{IJ} \Bigr) \eta_{J}  - \e^{-X}  \partial_{[a} X F^{(3)}_{\;\;bcd]} \\
& \quad - \frac{1}{48} \e^{-(\phi+X)} H_{efg} \bar{\epsilon}_{I} \biggl( \left(\gamma_{bcda}^{\quad\;\; efg} + 36 \gamma_{[bc}^{\;\;\;\;e} \delta_{d}^{f} \delta_{a]}^{g} \right) \left( S^{(2)}_{0} \sigma^{3} + S^{(2)}_{3} \mathbb{I} \right) \\
& \qquad\qquad\qquad + 48 \gamma_{[b} \delta_{c}^{e} \delta^{f}_{d} \delta^{g}_{a]}  \left( S^{(1)}_{0} \mathbb{I} + \left(S^{(1)}_{1} - S^{(2)}_{2} \right) \sigma^{1}  + \left(S^{(1)}_{2} - S^{(2)}_{1} \right) i \sigma^{2} + S^{(1)}_{3} \sigma^{3} \right) \biggr) \eta_{J} M_{IJ} \\ 
& \qquad + \frac{1}{2} \e^{-X} F^{(1)}_{\;\;[a} \bar{\epsilon}_{I} \gamma_{bcd]} \left(S^{(2)}_{3} \sigma^{1} - S^{(2)}_{1} \sigma^{3} \right) \eta_{J} M_{IJ}\\
& \qquad\quad - \frac{1}{48} \e^{-X}  F^{(3)}_{\;\;efg} \bar{\epsilon}_{I} \biggl( \left(\gamma_{bcda}^{\quad\;\; efg} + 36 \gamma_{[bc}^{\;\;\;\;e} \delta_{d}^{f} \delta_{a]}^{g} \right) \left( S^{(2)}_{0} \sigma^{1} + S^{(2)}_{1} \mathbb{I} \right) \\
& \qquad\qquad\qquad\qquad\qquad\qquad\qquad\qquad + 48 \gamma_{[b} \delta_{c}^{e} \delta^{f}_{d} \delta^{g}_{a]}  \left(  S^{(2)}_{2}  \sigma^{3} + S^{(2)}_{3} i \sigma^{2} \right) \biggr) \eta_{J} M_{IJ} \\ 
& \qquad\qquad - \frac{1}{4} \e^{-X} F^{(5)}_{\;\;ef[bcd} \bar{\epsilon}_{I} \left( \gamma^{e} \delta_{a]}^{f} \left(S^{(2)}_{0} i \sigma^{2} - S^{(2)}_{2} \mathbb{I} \right) + \gamma_{a]}^{\;\;ef} \left(S^{(2)}_{1} \sigma^{3} - S^{(2)}_{3} \sigma^{1} \right)\right) \eta_{J} M_{IJ},
\end{align*}
\begin{align*}
&\nabla_{[a} F'^{(5)}_{\;\;\;\;bcdef]} + \frac{10}{3} F'^{(3)}_{\;\;\;\;[abc} H_{def]}  \\ 
= &\e^{-(\phi+X)} \bar{\epsilon}_I \gamma_{[bcdef} \Bigl( \partial_{a]}  \left(S^{(3)} M_{IJ} \right) - \partial_{a]} X  S^{(3)} M_{IJ} \Bigr) \eta_{J}  - \e^{-X}  \partial_{[a} X F^{(5)}_{\;\;bcdef]} \\
& \quad - \frac{1}{72} \e^{-(\phi+X)} H_{ghi} \bar{\epsilon}_{I} \biggl( \left(\gamma_{bcdefa}^{\qquad \; ghi} + 90 \gamma_{[bcde}^{\quad\;\;\; g} \delta_{f}^{h} \delta_{a]}^{i} \right) \left( S^{(3)}_{0} \sigma^{3} + S^{(3)}_{3} \mathbb{I} \right) \\
& \qquad\qquad\quad + 240 \gamma_{[bcd} \delta_{e}^{g} \delta^{h}_{f} \delta^{i}_{a]}  \left( S^{(2)}_{0} \mathbb{I} + \left(S^{(2)}_{1} - S^{(3)}_{2} \right) \sigma^{1} + \left(S^{(2)}_{2} - S^{(3)}_{1} \right) i \sigma^{2} + S^{(2)}_{3} \sigma^{3} \right) \biggr) \eta_{J} M_{IJ}\\ 
& \qquad - \frac{1}{12} \e^{-X} F^{(1)}_{\;\;g} \bar{\epsilon}_{I} \gamma_{bcdefa}^{\qquad \; g} \left(S^{(3)}_{0} i \sigma^{2} - S^{(3)}_{2} \mathbb{I} \right) \eta_{J} M_{IJ}\\
& \qquad\quad - \frac{1}{72} \e^{-X}  F^{(3)}_{\;\;ghi} \bar{\epsilon}_{I} \biggl(\gamma_{bcdefa}^{\qquad \; ghi} \left( S^{(3)}_{0} \sigma^{1} + S^{(3)}_{1} \mathbb{I} \right) \\
& \qquad\qquad\qquad\qquad\qquad + \left(9 \gamma_{[bcdef}^{\qquad gh} \delta_{a]}^{i} + 60 \gamma_{[bcd} \delta_{e}^{g} \delta^{h}_{f} \delta^{i}_{a]} \right) \left(  S^{(3)}_{2}  \sigma^{3} + S^{(3)}_{3} i \sigma^{2} \right) \biggr) \eta_{J} M_{IJ} \\ 
& \qquad\qquad + \frac{1}{4} \e^{-X} F^{(5)}_{\;\;g[defa} \bar{\epsilon}_{I} \left(5 \gamma_{bc]}^{\;\;\; g} \left(S^{(3)}_{0} i \sigma^{2} - S^{(3)}_{2} \mathbb{I} \right) - 4 \gamma_{b} \delta_{c]}^{g} \left(S^{(3)}_{1} \sigma^{3} - S^{(3)}_{3} \sigma^{1} \right)\right) \eta_{J} M_{IJ}.
\end{align*}
Both of the above expressions must vanish for the Bianchi identities for $F'^{(3)}$ and $F'^{(5)},$ respectively, to hold. So we have that 
\begin{gather}
\partial_{a} X = \bar{\epsilon}_{I} \gamma_{a} \left(S^{(2)}_{2} \sigma^{3} + S^{(2)}_{3}  i \sigma^{2} \right) \eta_{J} M_{IJ}, \label{eqn:bian3condX}\\
\bar{\epsilon}_{I} \gamma_{abc} \left( S^{(2)}_{0} \sigma^{3} + S^{(2)}_{3} \mathbb{I} \right) \eta_{J} M_{IJ}=0, \label{eqn:bian3conG3S0S3} \\
\bar{\epsilon}_{I} \gamma_{a} \left( S^{(1)}_{0} \mathbb{I} + \left(S^{(1)}_{1} - S^{(2)}_{2} \right) \sigma^{1} 
+ \left(S^{(1)}_{2} - S^{(2)}_{1} \right) i \sigma^{2} + S^{(1)}_{3} \sigma^{3} \right) \eta_{J} M_{IJ}=0, \label{eqn:bian3conG1Sall}\\
\bar{\epsilon}_{I} \gamma_{abc} \left(S^{(2)}_{3} \sigma^{1} - S^{(2)}_{1} \sigma^{3} \right)\eta_{J} M_{IJ}=0, \label{eqn:bian3conG3S1S3}\\
\bar{\epsilon}_{I} \gamma_{abc} \left( S^{(2)}_{0} \sigma^{1} + S^{(2)}_{1} \mathbb{I} \right) \eta_{J} M_{IJ}=0, \\
\bar{\epsilon}_{I} \gamma_{a} \left(S^{(2)}_{0} i \sigma^{2} - S^{(2)}_{2} \mathbb{I} \right) \eta_{J} M_{IJ}=0, \\
\bar{\epsilon}_I \gamma_{[bcd} \Bigl( \partial_{a]}  \left(S^{(2)} M_{IJ} \right) - \partial_{a]} X  S^{(2)} M_{IJ} \Bigr) \eta_{J}=0,
\end{gather}
and
\begin{gather}
\partial_{a} X = \bar{\epsilon}_{I} \gamma_{a} \left(S^{(3)}_{1} \sigma^{3} - S^{(3)}_{3}  \sigma^{1} \right) \eta_{J} M_{IJ}, \\
\bar{\epsilon}_{I} \gamma_{a} \left( S^{(3)}_{0} \sigma^{3} + S^{(3)}_{3} \mathbb{I} \right) \eta_{J} M_{IJ}=0, \\
\bar{\epsilon}_{I} \gamma_{abcde} \left( S^{(3)}_{0} \sigma^{3} + S^{(3)}_{3} \mathbb{I} \right) \eta_{J} M_{IJ}=0, \\
\bar{\epsilon} \gamma_{abc} \left( S^{(2)}_{0} \mathbb{I} + \left(S^{(2)}_{1} - S^{(3)}_{2} \right) \sigma^{1} + \left(S^{(2)}_{2} - S^{(3)}_{1} \right) i \sigma^{2} + S^{(2)}_{3} \sigma^{3} \right) \eta_{J} M_{IJ}=0, \\
\bar{\epsilon}_{I} \gamma_{abc}\left(S^{(3)}_{0} i \sigma^{2} - S^{(3)}_{2} \mathbb{I} \right) \eta_{J} M_{IJ}=0, \\
\bar{\epsilon}_{I} \gamma_{a} \left(S^{(3)}_{0} \sigma^{1} + S^{(3)}_{1} \mathbb{I} \right) \eta_{J} M_{IJ}=0, \\
\bar{\epsilon}_{I} \gamma_{abc}  \left(S^{(3)}_{2} \sigma^{3} + S^{(3)}_{3} i \sigma^{2} \right) \eta_{J} M_{IJ}=0, \\
\bar{\epsilon}_{I} \gamma_{abc} \left(S^{(3)}_{0} i \sigma^{2} - S^{(3)}_{2} \mathbb{I} \right) \eta_{J} M_{IJ}=0, \\
\bar{\epsilon}_I \gamma_{[bcdef} \Bigl( \partial_{a]}  \left(S^{(3)} M_{IJ} \right) - \partial_{a]} X  S^{(3)} M_{IJ} \Bigr) \eta_{J}=0, \label{eqn:bian5conquart}
\end{gather}
respectively.

The NSNS 3-form field strength does not change, so the Bianchi identity for the 3-form field is the same as before.

Now, we assume that equations (\ref{eqn:bian1condX}--\ref{eqn:bian5conquart}) hold and consider the equations of motion. As before, using the Killing spinor equations for $\epsilon_{I}$ and $\eta_{J},$ the equation of motion for the transformed RR 1-form field strength can be simplified to  
\begin{align*}
&\nabla^{a} F'^{(1)}_{\;\;\;\;a} + \frac{1}{6} H_{abc} F'^{(3) abc}\\
= & \e^{-(\phi+X)} \bar{\epsilon}_{I} \gamma^{a} \Bigl(\partial_{a} \left(S^{(1)}   M_{IJ}\right)  - \partial_{a} X S^{(1)} M_{IJ} \Bigr) \eta_{J}  \\
& \qquad +  \frac{1}{6} \e^{-(\phi+X)} H_{abc} \bar{\epsilon}_{I} \gamma^{abc}  \left( S^{(2)}_{0} \mathbb{I}+ \left(S^{(2)}_{1} - S^{(1)}_{2} \right) \sigma^{1} + \left(S^{(2)}_{2} - S^{(1)}_{1} \right) i \sigma^{2} + S^{(2)}_{3} \sigma^{3} \right) \eta_{J} M_{IJ}=0.
\end{align*}
So, we get the following conditions:
\begin{gather}
\bar{\epsilon}_{I} \gamma_{abc}  \left( S^{(2)}_{0} \mathbb{I} + \left(S^{(2)}_{1} - S^{(1)}_{2} \right) \sigma^{1} + \left(S^{(2)}_{2} - S^{(1)}_{1} \right) i \sigma^{2} + S^{(2)}_{3} \sigma^{3} \right) \eta_{J} M_{IJ}=0, \\
\bar{\epsilon}_{I} \gamma^{a} \Bigl(\partial_{a} \left(S^{(1)}   M_{IJ}\right)  - \partial_{a} X S^{(1)} M_{IJ} \Bigr) \eta_{J}=0.
\label{conE1}
\end{gather}

Similarly, the equation of motion for the transformed 3-form field strength becomes
\begin{align*}
&\nabla^{a} F'^{(3)}_{\;\;\;\;abc} + \frac{1}{6} H^{def} F'^{(5)}_{\quad\; bcdef}\\
= & \e^{-(\phi+X)} \bar{\epsilon}_{I} \gamma_{abc} \Bigl(\partial^{a} \left(S^{(2)}   M_{IJ}\right)  - \partial^{a} X S^{(2)} M_{IJ} \Bigr) \eta_{J}  \\
& \quad + \frac{1}{6} \e^{-(\phi+X)} H^{def} \bar{\epsilon}_{I} \gamma_{bcdef}  \left( S^{(3)}_{0} \mathbb{I} + \left(S^{(3)}_{1} - S^{(2)}_{2} \right) \sigma^{1} + \left(S^{(3)}_{2} - S^{(2)}_{1} \right) i \sigma^{2} + S^{(3)}_{3} \sigma^{3} \right) \eta_{J} M_{IJ}=0.
\end{align*}
hence we need to impose
\begin{gather}
\bar{\epsilon}_{I} \gamma_{abcde}  \left( S^{(3)}_{0} \mathbb{I} + \left(S^{(3)}_{1} - S^{(2)}_{2} \right) \sigma^{1} + \left(S^{(3)}_{2} - S^{(2)}_{1} \right) i \sigma^{2} + S^{(3)}_{3} \sigma^{3} \right) \eta_{J} M_{IJ}=0, \label{eqn:eom3conG5} \\
\bar{\epsilon}_{I} \gamma_{abc} \Bigl(\partial^{a} \left(S^{(2)}   M_{IJ}\right)  - \partial^{a} X S^{(2)} M_{IJ} \Bigr) \eta_{J}=0.
\label{conE3}
\end{gather}

We also need to show that the transformed fields satisfy the equation of motion for the NSNS 3-form:
\begin{align}
&\nabla^{a} \left( \e^{-2\phi'} H_{abc} \right) - F'^{(1)a} F'^{(3)}_{\;\;abc} - \frac{1}{6} F'^{(3)def} F'^{(5)}_{\;\;bcdef} \notag \\
= & -2 \e^{-2(\phi +X)} \partial^{a} X H_{abc} - \e^{-(\phi + 2 X)} M_{IJ} \left( F^{(1)a} \bar{\epsilon}_{I} \gamma_{abc} S^{(2)} \eta_{J} + F^{(3)}_{\;\;\;abc} \bar{\epsilon}_{I} \gamma^{a} S^{(1)} \eta_{J} \right.\notag \\
& \qquad \quad\left. + \frac{1}{6} F^{(3)def} \bar{\epsilon}_{I} \gamma_{bcdef} S^{(3)} \eta_{J} +\frac{1}{6}  F^{(5)}_{\;\;bcdef} \bar{\epsilon}_{I} \gamma^{def} S^{(2)} \eta_{J} \right) \notag \\
& \qquad \qquad \qquad- \e^{-2(\phi + X)} \left( \bar{\epsilon}_{I} \gamma^{a} S^{(1)} \eta_{J} \right) \left( \bar{\epsilon}_{K} \gamma_{abc} S^{(2)} \eta_{L} \right) M_{IJ} M_{KL} \notag \\
& \qquad \qquad \qquad\qquad\qquad- \frac{1}{6} \e^{-2(\phi + X)} \left( \bar{\epsilon}_{I} \gamma^{def} S^{(2)} \eta_{J} \right) \left( \bar{\epsilon}_{K} \gamma_{bcdef} S^{(3)} \eta_{L} \right) M_{IJ} M_{KL}=0,
\label{eqn:transHeqnB}
\end{align}
where the NSNS 3-form equation of motion with the original supergravity fields has been used in the first equality. 

Using the gravitino Killing spinor equation and the self-duality of the 5-form RR field strength we can show that 
\begin{align}
F^{(1)a} \bar{\epsilon}_{I} \gamma_{abc} S^{(2)}_{2} i\sigma^{2} \eta_{J} M_{IJ} = & 4 \nabla_{[b} \left( \bar{\epsilon}_{I} \gamma_{c]} \eta_{J} \right) S^{(2)}_{2} M_{IJ} - 2 H_{abc} \bar{\epsilon}_{I} \gamma^{a} S^{(2)}_{2} \sigma^{3} \eta_{J} M_{IJ} \notag \\
& \qquad - \frac{1}{6} \e^{\phi} F^{(3)}_{\;\;\;def} \bar{\epsilon}_{I} \left( \gamma_{bc}^{\;\;\;def} + \gamma^{d} \delta_{b}^{e} \delta_{c}^{f} \right) S^{(2)}_{2} \sigma^{1} \eta_{J} M_{IJ} \notag \\
& \qquad \qquad\qquad - \frac{1}{6} \e^{\phi} F^{(5)}_{\;\;\;bcdef} \bar{\epsilon}_{I} \gamma^{def} S^{(2)}_{2} i \sigma^{2} \eta_{J} M_{IJ},
\end{align}
and using the dilatino Killing spinor equation and equations (\ref{eqn:bian3conG3S0S3}--\ref{eqn:bian3conG3S1S3}) we get that 
\begin{align}
&F^{(1)a} \bar{\epsilon}_{I} \gamma_{abc} \left( S^{(2)}_{0} \mathbb{I} + S^{(2)}_{1} \sigma^{1} + S^{(2)}_{3} \sigma^{3} \right) \eta_{J} M_{IJ} \notag \\
= &- 2 \partial_{[b} \phi \bar{\epsilon}_{I} \gamma_{c]} S^{(2)}_{0} i \sigma^{2} \eta_{J} M_{IJ} + \frac{1}{12} H_{def} \bar{\epsilon}_{I} \left( \gamma_{bc}^{\;\;\;def} - 6 \gamma^{d} \delta_{b}^{e} \delta_{c}^{f} \right) S^{(2)}_{3} i \sigma^{2} \eta_{J} M_{IJ} \notag \\
& \qquad\qquad\qquad\qquad\qquad\quad\qquad + \frac{1}{12} \e^{\phi} F^{(3)}_{\;\;\;def} \bar{\epsilon}_{I} \left( \gamma_{bc}^{\;\;\;def} - 6 \gamma^{d} \delta_{b}^{e} \delta_{c}^{f} \right) S^{(2)}_{1} i \sigma^{2} \eta_{J} M_{IJ}=0.
\end{align}
Substituting the above equations into equation \eqref{eqn:transHeqnB}, and using equations \eqref{eqn:bian3condX}, \eqref{eqn:bian3conG1Sall} and \eqref{eqn:eom3conG5}, the NSNS 3-form equation of motion becomes 
\begin{align}
& -4 \nabla_{[b} \left( \bar{\epsilon}_{I} \gamma_{c]} \eta_{J} \right) S^{(2)}_{2} M_{IJ} + 2  \partial_{[b} \phi \bar{\epsilon}_{I} \gamma_{c]} S^{(2)}_{0} i \sigma^{2} \eta_{J} M_{IJ} - \frac{1}{12} H_{def} \bar{\epsilon}_{I} \left( \gamma_{bc}^{\;\;\;def} + 18 \gamma^{d} \delta_{b}^{e} \delta_{c}^{f} \right) S^{(2)}_{3} i \sigma^{2} \eta_{J} M_{IJ}\notag \\
& \; - \frac{1}{4} \e^{\phi} F^{(3)}_{\;\;\;def} \bar{\epsilon}_{I}  \left( \gamma_{bc}^{\;\;\;def} + 2 \gamma^{d} \delta_{b}^{e} \delta_{c}^{f} \right) S^{(2)}_{1} i \sigma^{2} \eta_{J} M_{IJ} - \frac{1}{6} \e^{\phi} F^{(5)}_{\;\;\;bcdef} \bar{\epsilon}_{I} \gamma^{def} \left( S^{(2)}_{0} \mathbb{I} + S^{(2)}_{1} \sigma^{1} + S^{(2)}_{3} \sigma^{3} \right) \eta_{J} M_{IJ}\notag  \\
& \;\;\; -  \left( \bar{\epsilon}_{I} \gamma^{a} S^{(1)} \eta_{J} \right) \left( \bar{\epsilon}_{K} \gamma_{abc} S^{(2)} \eta_{L} \right) M_{IJ} M_{KL}- \frac{1}{6} \left( \bar{\epsilon}_{I} \gamma^{def} S^{(2)} \eta_{J} \right) \left( \bar{\epsilon}_{K} \gamma_{bcdef} S^{(3)} \eta_{L} \right) M_{IJ} M_{KL}=0.
\label{eqn:transHeqn2B}
\end{align}
The supergravity fields are generic so the terms proportional to each supergravity field in the above expression must vanish. In particular, if we consider the expression proportional to the RR 5-form field strength, then as this expression is exactly the expression that enters in the transformation of the RR 3-form field strength $$S^{(2)}_{0} = S^{(2)}_{1} = S^{(2)}_{3}=0,$$ and without loss of generality we can set $$S^{(2)}_2=1.$$

Similarly, by using the Killing spinor equations to substitute in for $$ F^{(3)}_{\;\;\;abc} \bar{\epsilon}_{I} \gamma^{a} S^{(1)} \eta_{J} \quad \textup{and} \quad F^{(3)}_{\;\;\;def} \bar{\epsilon}_{I} \gamma_{bc}^{\;\;\;def} S^{(3)} \eta_{J} $$ in equation \eqref{eqn:transHeqnB}, instead of $F^{(1)}_{\;\;\;a} \bar{\epsilon}_{I} \gamma^{a}_{\;\,bc} S^{(2)} \eta_{J}$, we can show that $$ S^{(1)}= \sigma^{1} \quad \textup{and} \quad S^{(3)}= \sigma^{1}.$$ 

Letting $$ S^{(1)}= S^{(3)}= \sigma^{1} \quad \textup{and} \quad S^{(2)}= i \sigma^{2},$$ conditions (\ref{eqn:bian1condX}--\ref{conE3}) become 
\begin{gather}
\partial_{a} X = \bar{\epsilon}_{I} \gamma_{a} \sigma^{3} \eta_{J} M_{IJ}, \label{eqn:condXB} \\
\bar{\epsilon}_{I} \gamma_{a}\eta_{J} M_{IJ}=0, \label{eqn:congamma1} \\
\bar{\epsilon}_I \gamma_{[b} \sigma^{1} \eta_{J}  \left( \partial_{a]} M_{IJ} - \partial_{a]} X  M_{IJ} \right) =0, \label{eqn:B1conquart} \\
\bar{\epsilon}_I \gamma_{[bcd} i \sigma^{2} \eta_{J} \left( \partial_{a]} M_{IJ} - \partial_{a]} X  M_{IJ} \right) =0, \\
\bar{\epsilon}_I \gamma_{[bcdef} \sigma^{1} \eta_{J} \left( \partial_{a]}  M_{IJ} - \partial_{a]} X  M_{IJ} \right) =0, \\
\bar{\epsilon}_{I} \gamma^{a} \sigma^{1} \eta_{J} \left(\partial_{a} M_{IJ} - \partial_{a} X M_{IJ} \right) =0, \\
\bar{\epsilon}_{I} \gamma_{abc} i \sigma^{2} \eta_{J} \left(\partial^{a} M_{IJ} - \partial^{a} X M_{IJ} \right) =0. \label{eqn:E3conquart}
\end{gather}
The NSNS 3-form field strength equation of motion, equation \eqref{eqn:transHeqn2B}, becomes
\begin{align*}
& 4 \nabla_{[b} \left( \bar{\epsilon}_{I} \gamma_{c]} \eta_{J} \right) M_{IJ} + \left( \bar{\epsilon}_{I} \gamma^{a} \sigma^{1} \eta_{J} \right) \left( \bar{\epsilon}_{K} \gamma_{abc} i \sigma^{2} \eta_{L} \right) M_{IJ} M_{KL} \notag \\
& \qquad \qquad \qquad\qquad\qquad\qquad\qquad\qquad+ \frac{1}{6} \left( \bar{\epsilon}_{I} \gamma^{def} i \sigma^{2} \eta_{J} \right) \left( \bar{\epsilon}_{K} \gamma_{bcdef} \sigma^{1} \eta_{L} \right) M_{IJ} M_{KL}=0,
\end{align*}
and using equation \eqref{eqn:congamma1} this reduces to  
\begin{align}
& 4 \bar{\epsilon}_{I} \gamma_{[b} \eta_{J} \partial_{c]} M_{IJ} + \left( \left( \bar{\epsilon}_{I} \gamma^{a} \sigma^{1} \eta_{J} \right) \left( \bar{\epsilon}_{K} \gamma_{abc} i \sigma^{2} \eta_{L} \right)+ \frac{1}{6} \left( \bar{\epsilon}_{I} \gamma^{def} i \sigma^{2} \eta_{J} \right) \left( \bar{\epsilon}_{K} \gamma_{bcdef} \sigma^{1} \eta_{L} \right) \right)M_{IJ} M_{KL}=0. \notag \\
& \qquad \qquad \qquad\qquad\quad\;\;
\label{eqn:transHeqn3B}
\end{align}

Fierz identities can be used to simplify the terms that are quartic in spinors. Just as the tensor product of a combination of gamma matrices, $M$ and $N,$ can be expanded in the basis $\{\mathcal{O}_I\}= \{ \mathbb{I}, \gamma_a, i \gamma_{ab}, i \gamma_{abc}, \gamma_{abcd}, \dots\},$
\[
M^{\alpha}_{\;\; \beta} N^{\gamma}_{\;\; \delta}  = 2^{-[d/2]} \sum_{I} \left( M \mathcal{O}^I N \right)^{\alpha}_{\;\; \delta}  \mathcal{O}^{\; \gamma}_{I \;\; \beta},
\]
we can expand the tensor product of $2 \times 2$ matrices, $\Sigma$ and $\Xi$, in the basis $\sigma^{\mu}=(\mathbb{I}, \sigma^{1}, \sigma^{2}, \sigma^{3}),$ 
\[
\Sigma_{AB} \Xi_{CD}  = 2^{-1} \sum_{\mu} \left( \Sigma \sigma^{\mu} \Xi \right)_{AD} \sigma^{\mu}_{CB},
\]
where uppercase Latin letters are $SO(2)$ vector indices. Hence, for type IIB theory spinors, the Fierz identity is 
\[
\left( \bar{\lambda} M \Sigma \chi \right) \left( \bar{\psi} N \Xi \varphi \right) = \frac{1}{64} \sum_{I, \mu} \bar{\lambda} \left( M \mathcal{O}^I N\right) \left( \Sigma \sigma^{\mu} \Xi \right) \varphi \bar{\psi} \left(  \mathcal{O}_I \sigma^{\mu} \right) \chi.
\]

Using the Fierz identity multiple times, we can show that 
\begin{align*}
& \left( \bar{\epsilon}_{I} \gamma^{a} \sigma^1 \eta_{J} \right) \left(\bar{\epsilon}_{K} \gamma_{abc} i \sigma^2 \eta_{L} \right) + \frac{1}{6} \left(\bar{\epsilon}_{I} \gamma^{a_1 \dots a_3} i \sigma^2 \eta_{J} \right) \left(\bar{\epsilon}_{K}  \gamma_{a_1 \dots a_3 bc} \sigma^1 \eta_{L} \right) \\
= & -16 \left( \bar{\epsilon}_{I} \gamma_{[b} \sigma^3  \eta_{L} \right) \left(\bar{\epsilon}_{K} \gamma_{c]} \eta_{J} \right) + \left( \bar{\epsilon}_{I} \gamma^{a}  i \sigma^2  \eta_{J} \right) \left(\bar{\epsilon}_{K} \gamma_{abc} \sigma^1 \eta_{L} \right)+ \frac{1}{6} \left( \bar{\epsilon}_{I} \gamma^{a_1 \dots a_3} \sigma^1 \eta_{J} \right) \left(\bar{\epsilon}_{K}  \gamma_{a_1 \dots a_3 bc}  i \sigma^2 \eta_{L} \right).
\end{align*}

So from equation \eqref{eqn:transHeqn3B}, the NSNS 3-form equation of motion is satisfied if
\begin{align}
&  4 \bar{\epsilon}_{I} \gamma_{[b} \eta_{J} \left( \partial_{c]} M_{IJ} + 4 \bar{\epsilon}_{K} \gamma_{c]} \sigma^3  \eta_{L} M_{KJ} M_{IL} \right) + \left(\bar{\epsilon}_{I} \gamma^{a}  i \sigma^2  \eta_{J} \right) \left(\bar{\epsilon}_{K} \gamma_{abc} \sigma^1 \eta_{L} \right) M_{IJ} M_{KL}\notag \\
& \qquad \qquad \qquad \qquad \qquad \qquad \qquad  \qquad + \frac{1}{6} \left( \bar{\epsilon}_{I} \gamma^{a_1 \dots a_3} \sigma^1 \eta_{J} \right) \left(\bar{\epsilon}_{K}  \gamma_{a_1 \dots a_3 bc}  i \sigma^2 \eta_{L} \right) M_{IJ} M_{KL} =0.
\label{conEHB}
\end{align}

The dilaton equation, \eqref{dileom}, for the transformed fields simply reduces to
\begin{equation}
\Box X -2 \partial_{a} \phi \partial^{a} X - (\partial X)^{2} =0.
\label{condilB}
\end{equation}
Using $ \partial_{a} X = \bar{\epsilon}_{I} \gamma_{a} \sigma^{3} \eta_{J} M_{IJ},$ and the Killing spinor equations, the above equation reduces to 
\begin{equation*}
\bar{\epsilon}_{I} \gamma^{a} \sigma^3  \eta_{J} \left(\partial_{a} M_{IJ} - \bar{\epsilon}_{K} \gamma_{a} \sigma^3  \eta_{L} M_{IJ} M_{KL} \right)=0, 
\end{equation*}
which using Fierz identities becomes
\begin{align}
&\bar{\epsilon}_{I} \gamma^{a} \sigma^3  \eta_{J} \left(\partial_{a} M_{IJ} + 2 \bar{\epsilon}_{K} \gamma_{a} \sigma^3  \eta_{L} M_{IL} M_{KJ} \right) + \biggl( \left( \bar{\epsilon}_{I} \gamma_{a} \eta_{J} \right) \left( \bar{\epsilon}_{K} \gamma^{a}  \eta_{L} \right) + 2 \left( \bar{\epsilon}_{I} \gamma_{a} \eta_{L} \right) \left( \bar{\epsilon}_{K} \gamma^{a}  \eta_{J} \right)\notag \\
& \qquad \qquad\qquad \qquad  -\frac{1}{12} \left( \bar{\epsilon}_{I} \gamma_{abc} \eta_{J} \right) \left( \bar{\epsilon}_{K} \gamma^{abc}  \eta_{L} \right) -\frac{1}{12} \left( \bar{\epsilon}_{I} \gamma_{abc} \sigma^{3} \eta_{J} \right) \left( \bar{\epsilon}_{K} \gamma^{abc} \sigma^{3} \eta_{L} \right) \biggr) M_{IJ} M_{KL} =0.
 \label{dilred}
\end{align}
This must hold in order for the dilaton equation to be satisfied for the transformed fields.

Now, let us consider the Einstein equation. We can use the gravitino Killing spinor equation and the constraint from the dilaton equation of motion, equation \eqref{condilB}, to show that the Einstein equation reduces to
\begin{align*}
& 2 \nabla_{(a} \left( \bar{\epsilon}_{I} \gamma_{b)} \sigma^{3} \eta_{J} M_{IJ} - \partial_{b)} X \right) - \frac{1}{4} g_{ab} \nabla_{c} \left( \bar{\epsilon}_{I} \gamma^{c} \sigma^{3} \eta_{J} M_{IJ} - \partial^{c} X \right) - 2 \bar{\epsilon}_{I} \gamma_{(a} \sigma^{3} \eta_{J} \partial_{b)} M_{IJ}\\
& \quad  + \frac{1}{4} g_{ab} \bar{\epsilon}_{I} \gamma^{c} \sigma^{3} \eta_{J} \partial_{c} M_{IJ}+\frac{1}{96}  \biggl( 48 \left( \bar{\epsilon}_{I} \gamma_{a} \sigma^{1} \eta_{J} \right) \left(\bar{\epsilon}_{K} \gamma_{b} \sigma^{1} \eta_{L} \right) + 24  \left( \bar{\epsilon}_{I} \gamma_{a}^{\;\;cd} i \sigma^{2} \eta_{J} \right) \left(\bar{\epsilon}_{K} \gamma_{bcd} i \sigma^{2} \eta_{L} \right)\\
& \qquad \qquad  - 2 g_{ab} \left( \bar{\epsilon}_{I} \gamma_{cde} i \sigma^{2} \eta_{J} \right) \left(\bar{\epsilon}_{K} \gamma^{cde} \eta_{L} i \sigma^{2} \right) + \left( \bar{\epsilon}_{I} \gamma_{a}^{\;\;cdef} \sigma^{1} \eta_{J} \right) \left(\bar{\epsilon}_{K} \gamma_{bcdef} \sigma^{1} \eta_{L} \right) \biggr) M_{IJ} M_{KL}=0.  
\end{align*}
The first two terms vanish because of equation \eqref{eqn:condXB}, and we can use Fierz identities to rewrite the terms that are quartic in spinors. Upon doing so, Einstein's equation becomes 
\begin{align}
& - 2 \bar{\epsilon}_{I} \gamma_{(a} \sigma^{3} \epsilon_{J} \left( \partial_{b)} M_{IJ} + 2 \bar{\epsilon}_{K} \gamma_{b)} \sigma^3  \epsilon_{L} M_{IL} M_{KJ} \right) + \frac{1}{4} g_{ab} \bar{\epsilon}_{I} \gamma^{c} \sigma^{3} \epsilon_{J} \left( \partial_{c} M_{IJ} +2 \bar{\epsilon}_{K} \gamma_{c} \sigma^3  \epsilon_{L} M_{IL} M_{KJ} \right)\notag \\
& \qquad \qquad+\frac{1}{96}  \biggl(384 \left( \bar{\epsilon}_{I} \gamma_{a} \eta_{L} \right) \left(\bar{\epsilon}_{K} \gamma_{b} \eta_{J} \right) + 48 \left( \bar{\epsilon}_{I} \gamma_{a} i \sigma^{2} \eta_{J} \right) \left(\bar{\epsilon}_{K} \gamma_{b} i \sigma^{2} \eta_{L} \right) \notag\\
& \qquad \qquad\qquad\quad + 24  \left( \bar{\epsilon}_{I} \gamma_{a}^{\;\;cd} \sigma^{1} \eta_{J} \right) \left(\bar{\epsilon}_{K} \gamma_{bcd} \sigma^{1} \eta_{L} \right)+  \left( \bar{\epsilon}_{I} \gamma_{a}^{\;\;cdef} i \sigma^{2} \eta_{J} \right) \left(\bar{\epsilon}_{K} \gamma_{bcdef} i \sigma^{2}  \eta_{L} \right)  \notag\\
& \qquad \qquad\qquad\qquad \quad- 48 g_{ab} \left( \bar{\epsilon}_{I} \gamma_{c} \eta_{J} \right) \left(\bar{\epsilon}_{K} \gamma^{c} \eta_{L} \right) - 2 g_{ab} \left( \bar{\epsilon}_{I} \gamma_{cde} \sigma^{1} \eta_{J} \right) \left(\bar{\epsilon}_{K} \gamma^{cde} \sigma^{1} \eta_{L} \right)  \biggr) M_{IJ} M_{KL}=0. 
\label{Einscon}
\end{align}

So, the transformed fields satisfy the type IIB supergravity equations if equations (\ref{eqn:condXB}--\ref{eqn:E3conquart}), \eqref{conEHB}, \eqref{dilred} and \eqref{Einscon} are satisfied. Using Fierz identities, equations (\ref{eqn:B1conquart}--\ref{eqn:E3conquart}) are equivalent to
\begin{align}
&\bar{\epsilon}_{I} \gamma_{[a} \sigma^{1} \epsilon_{J} \left( \partial_{b]} M_{IJ} + 2 \bar{\epsilon}_{K} \gamma_{b]} \sigma^3  \epsilon_{L} M_{IL} M_{KJ} \right) - \biggl( \frac{1}{4} \left( \bar{\epsilon}_{I} \gamma_{abc} \eta_{J} \right) \left(\bar{\epsilon}_{K} \gamma^{c} i \sigma^{2} \eta_{L} \right) \notag \\
& \qquad\quad+ \frac{1}{2} \left( \bar{\epsilon}_{I} \gamma_{abc} i \sigma^{2} \eta_{J} \right) \left(\bar{\epsilon}_{K} \gamma^{c} \eta_{L} \right) + \left( \bar{\epsilon}_{I} \gamma_{abc} i \sigma^{2} \eta_{L} \right) \left(\bar{\epsilon}_{K} \gamma^{c} \eta_{J} \right)- \frac{1}{4} \left( \bar{\epsilon}_{I} \gamma_{[a}^{\;\;\;cd} \sigma^{3} \eta_{J} \right) \left(\bar{\epsilon}_{K} \gamma_{b]cd} \sigma^{1} \eta_{L} \right)\notag \\
& \qquad\qquad\qquad\qquad\qquad\qquad -\frac{1}{24} \left( \bar{\epsilon}_{I} \gamma_{abcde} i \sigma^{2} \eta_{J} \right) \left(\bar{\epsilon}_{K} \gamma^{cde}  \eta_{L} \right)  \biggr) M_{IJ} M_{KL}=0,
\label{eqn:B1conquart2}
\end{align}
\begin{align}
&\bar{\epsilon}_{I} \gamma_{[abc} i \sigma^{2} \epsilon_{J} \left( \partial_{d]} M_{IJ} + 2 \bar{\epsilon}_{K} \gamma_{d]} \sigma^3  \epsilon_{L} M_{IL} M_{KJ} \right) - \biggl( \frac{3}{4} \left( \bar{\epsilon}_{I} \gamma_{[ab}^{\;\;\;\;\, e} \sigma^{1} \eta_{J} \right) \left(\bar{\epsilon}_{K} \gamma_{cd]e} \eta_{L} \right)\notag \\
& \qquad \quad  + \frac{1}{2} \left( \bar{\epsilon}_{I} \gamma_{[abc} \sigma^{3} \eta_{J} \right) \left(\bar{\epsilon}_{K} \gamma_{d]} i \sigma^{2} \eta_{L} \right) + \frac{1}{4} \left( \bar{\epsilon}_{I} \gamma_{abcde} \sigma^{1} \eta_{J} \right) \left(\bar{\epsilon}_{K} \gamma^{e} \eta_{L} \right)  + \frac{1}{2} \left( \bar{\epsilon}_{I} \gamma_{abcde} \sigma^{1} \eta_{L} \right) \left(\bar{\epsilon}_{K} \gamma^{e} \eta_{J} \right)\notag\\
& \qquad \qquad\qquad  + \frac{1}{4} \left( \bar{\epsilon}_{I} \gamma_{[abc}^{\quad \;\, e f} i \sigma^{2} \eta_{J} \right) \left(\bar{\epsilon}_{K} \gamma_{d]ef} \sigma^{3} \eta_{L} \right)+ \frac{1}{48} \left( \bar{\epsilon}_{I} \gamma_{abcdefg}\eta_{J} \right) \left(\bar{\epsilon}_{K} \gamma^{efg} \sigma^{1} \eta_{L} \right)  \biggr) M_{IJ} M_{KL}=0,
\label{eqn:B3conquart2}
\end{align}
\begin{align}
&\bar{\epsilon}_{I} \gamma_{[abcde} \sigma^{1} \epsilon_{J} \left( \partial_{f]} M_{IJ} + 2 \bar{\epsilon}_{K} \gamma_{f]} \sigma^3  \epsilon_{L} M_{IL} M_{KJ} \right) - \biggl( \frac{5}{3} \left( \bar{\epsilon}_{I} \gamma_{[abc} \sigma^{3} \eta_{J} \right) \left(\bar{\epsilon}_{K} \gamma_{def]} \sigma^{1} \eta_{L} \right)\notag \\
& \qquad \quad  + \frac{5}{4} \left( \bar{\epsilon}_{I} \gamma_{[abcd}^{\quad \;\;\; g} i \sigma^{2} \eta_{J} \right) \left(\bar{\epsilon}_{K} \gamma_{ef]g} \eta_{L} \right) - \frac{1}{12} \left( \bar{\epsilon}_{I} \gamma_{abcdefg} \eta_{J} \right) \left(\bar{\epsilon}_{K} \gamma^{g} i \sigma^{2} \eta_{L} \right) \notag\\
& \qquad \qquad \quad \;\;\;+ \frac{1}{4} \left( \bar{\epsilon}_{I} \gamma_{[abcde}^{\qquad g h} \sigma^{1} \eta_{J} \right) \left(\bar{\epsilon}_{K} \gamma_{f]g h} \sigma^{3} \eta_{L} \right) + \frac{1}{6} \left( \bar{\epsilon}_{I} \gamma_{abcdefg} i \sigma^{2} \eta_{J} \right) \left(\bar{\epsilon}_{K} \gamma^{g}  \eta_{L} \right)\notag\\
& \qquad \qquad\qquad \qquad \quad+ \frac{1}{3} \left( \bar{\epsilon}_{I} \gamma_{abcdefg} i \sigma^{2} \eta_{L} \right) \left(\bar{\epsilon}_{K} \gamma^{g}  \eta_{J} \right)  \biggr) M_{IJ} M_{KL}=0,
\label{eqn:B5conquart2}
\end{align}
\begin{align}
&\bar{\epsilon}_{I} \gamma^{a} \sigma^{1} \epsilon_{J} \left( \partial_{a} M_{IJ} + 2 \bar{\epsilon}_{K} \gamma_{a} \sigma^3  \epsilon_{L} M_{IL} M_{KJ} \right)  -  \frac{1}{12} \left( \bar{\epsilon}_{I} \gamma_{abc} \sigma^{1} \eta_{J} \right) \left(\bar{\epsilon}_{K} \gamma^{abc} \sigma^{3} \eta_{L} \right) M_{IJ} M_{KL}=0,
\label{eqn:E1conquart2}
\end{align}
\begin{align}
&\bar{\epsilon}_{I} \gamma_{abc} i \sigma^{2} \epsilon_{J} \left( \partial^{c} M_{IJ} + 2 \bar{\epsilon}_{K} \gamma^{c} \sigma^3  \epsilon_{L} M_{IL} M_{KJ} \right) + \biggl( 2 \left( \bar{\epsilon}_{I} \gamma_{[a} \eta_{J} \right) \left(\bar{\epsilon}_{K} \gamma_{b]} \sigma^{1} \eta_{L} \right) \notag \\
& \qquad \quad+ 4 \left( \bar{\epsilon}_{I} \gamma_{[a} \eta_{L} \right) \left(\bar{\epsilon}_{K} \gamma_{b]} \sigma^{1} \eta_{J} \right)- \frac{1}{2} \left( \bar{\epsilon}_{I} \gamma_{[a}^{\;\;\;cd} \eta_{J} \right) \left(\bar{\epsilon}_{K} \gamma_{b]cd} \sigma^{1} \eta_{L} \right) + \frac{1}{2} \left( \bar{\epsilon}_{I} \gamma_{abc} \sigma^{3} \eta_{J} \right) \left(\bar{\epsilon}_{K} \gamma^{c} i \sigma^{2} \eta_{L} \right)\notag \\
& \qquad \qquad\qquad\qquad\qquad\qquad  - \frac{1}{12} \left( \bar{\epsilon}_{I} \gamma_{abcde} i \sigma^{2} \eta_{J} \right) \left(\bar{\epsilon}_{K} \gamma^{cde} \sigma^{3} \eta_{L} \right) \biggr) M_{IJ} M_{KL}=0,
\label{eqn:E3conquart2}
\end{align}
respectively.

A solution to equations \eqref{conEHB}, (\ref{dilred}--\ref{eqn:E3conquart2}) is 
\begin{gather*}
\bar{\epsilon}_{I} \gamma_{a}  \eta_{J}  =0, \quad \bar{\epsilon}_{I} \gamma_{a} i \sigma^{2} \eta_{J} M_{IJ}  =0,  \\ 
\bar{\epsilon}_{I} \gamma_{abc}  \eta_{J} M_{IJ} =0, \quad \bar{\epsilon}_{I} \gamma_{abc}  \sigma^{1} \eta_{J} M_{IJ}  =0,  \\
\bar{\epsilon}_{I} \gamma_{abc}  \sigma^{3} \eta_{J} M_{IJ}  =0, \quad \bar{\epsilon}_{I} \gamma_{abcde} i \sigma^{2} \eta_{J} M_{IJ}  =0, \\
\partial_{a} M_{IJ} = -2 \bar{\epsilon}_{K} \gamma_{a} \sigma^3  \eta_{L} M_{IL} M_{KJ}.
\end{gather*}

Therefore, the type IIB supergravity symmetry is described by the following transformations of the RR fields and dilaton 
\begin{align}
\phi \rightarrow \phi' &= \phi + X,\notag \\
\e^{\phi} F^{(1)}_{\;\;\;\;a} \rightarrow \e^{\phi'} F'^{(1)}_{\;\;\;\;a} &= \e^{\phi} F^{(1)}_{\;\;\;\;a} + \bar{\epsilon}_{I} \gamma_{a} \sigma^{1} \eta_{J} M_{IJ}, \notag \\
\e^{\phi} F^{(3)}_{\;\;\;\;abc} \rightarrow \e^{\phi'} F'^{(3)}_{\;\;\;\;abc} &= \e^{\phi} F^{(3)}_{\;\;\;\;abc} + \bar{\epsilon}_{I} \gamma_{abc} i \sigma^{2} \eta_{J} M_{IJ}, \notag \\
\e^{\phi} F^{(5)}_{\;\;\;\;abcde} \rightarrow \e^{\phi'} F'^{(5)}_{\;\;\;\;abcde} &= \e^{\phi} F^{(5)}_{\;\;\;\;abcde} + \bar{\epsilon}_{I} \gamma_{abcde} \sigma^{1} \eta_{J} M_{IJ},
\label{eqn:IIBsymm}
\end{align}
where 
\begin{gather}
\gamma_{11} \epsilon_{I} = \epsilon_{I}, \quad \gamma_{11} \eta_{I} = \eta_{I}, \\
 \bar{\epsilon}_{I} \gamma_{a}  \eta_{J}  =0,   \label{eqn2:congamma1} \\ 
\bar{\epsilon}_{I} \gamma_{a} i \sigma^{2} \eta_{J} M_{IJ}  =0,\quad \bar{\epsilon}_{I} \gamma_{abc}  \eta_{J} M_{IJ} =0, \quad \bar{\epsilon}_{I} \gamma_{abc}  \sigma^{1} \eta_{J} M_{IJ}  =0,  \label{con2B} \\
\bar{\epsilon}_{I} \gamma_{abc}  \sigma^{3} \eta_{J} M_{IJ}  =0, \quad \bar{\epsilon}_{I} \gamma_{abcde} i \sigma^{2} \eta_{J} M_{IJ}  =0,  \label{con3B} \\
\partial_{a} X  = \bar{\epsilon}_{I} \gamma_{a} \sigma^{3} \eta_{J} M_{IJ},  \label{XdefB} \\
\partial_{a} M_{IJ} = -2 \bar{\epsilon}_{K} \gamma_{a} \sigma^3  \eta_{L} M_{IL} M_{KJ}.
\label{eqn:MdefB}
\end{gather}
Equation \eqref{eqn:MdefB} is equivalent to 
\begin{equation}
\partial_{a} (M^{-1})_{IJ} =  2 \bar{\epsilon}_{J} \gamma_{a} \sigma^3 \eta_{I},
\end{equation}
and up to a constant of integration
\begin{equation}
X =  \frac{1}{2}  \sum_{I=1}^{n}\left(\log M^{-1} \right)_{II}.
\end{equation} 

The integrability conditions for equations \eqref{XdefB} and \eqref{eqn:MdefB} are satisfied by equation \eqref{eqn2:congamma1}.

If  $\epsilon_{I} = \eta_{I}$ then the equations in the lines labelled by \eqref{con2B} and \eqref{con3B} are satisfied, and the transformations are precisely the transformations found by Berkovits and Maldacena, equations \eqref{eqn:BMFtrans} and \eqref{eqn:BMdiltrans} in section \ref{review}. Furthermore, when $n=1$ these equations can be explicitly solved to show that $\epsilon \propto \eta.$ When $n>1$ this is no longer true, and the conditions can be satisfied without identifying $\epsilon_{I}$ and $\eta_{I}.$

Note that, since in our transformations it is not necessary to identify $\epsilon_{I}$ and $\eta_{I},$ we can solve $$\bar{\epsilon}_{I} \gamma_{a}  \eta_{J}  =0$$ for real spinors.

\section{Comments} 
\label{com}

In both type IIA and type IIB supergravity we have found a larger class of transformations that include the transformations of Berkovits and Maldacena \cite{fermdual}. In both cases, when $n=1,$ these transformations are precisely the transformations found by Berkovits and Maldacena. However, for $n>1$ $$\epsilon_{I} \propto \eta_{I}$$ is sufficient but no longer necessary for the conditions given by equations (\ref{con2A1}, \ref{con2A2}) and (\ref{con2B}, \ref{con3B}) in the analysis for type IIA and type IIB supergravity, respectively, to be satisfied. Indeed, in both cases, we have found spinors $\epsilon_{I} \neq \eta_{I},$ where $I=1,2,$ for which $M_{IJ}$ is antisymmetric in its $I,J$ indices and the above-mentioned conditions hold.

In the transformations of fermionic T-duality, the spinors were complexified in order to find non-trivial solutions to \begin{equation*}
\bar{\epsilon}_{I} \gamma_{a} \epsilon_{J}=0.
\end{equation*}
Note that, in the transformations that we have constructed $\eta_{I}$ does not necessarily have to be identified with $\epsilon_{I}$ when $n>1.$ Therefore, 
\begin{equation*}
\bar{\epsilon}_{I} \gamma_{a} \eta_{J}=0
\end{equation*}
can be solved for real spinors, keeping the transformation real.

Furthermore, when the two set of Killing spinors $\epsilon_I$ and $\eta_I$ are identified the supersymmetry of the transformed supergravity solution is the same as the original solution. In fact, the Killing spinors in the new background can be written explicitly in terms of the Killing spinors of the original background \cite{fermdual}, equation \eqref{newkillingspinors}. This must be true because fermionic T-duality is a duality of string theory, so the transformation must preserve supersymmetry. However, for our transformation it is not clear whether supersymmetry is preserved. If this is case, then the transformation could be a useful tool for generating backgrounds with lower supersymmetry.

In general, however, the conditions given in equations (\ref{eqn2:gamma10con}--\ref{con2A2}) and (\ref{eqn2:congamma1}--\ref{con3B}) are difficult to solve explicitly. If this symmetry, and indeed fermionic T-duality, is to be a more practical solution-generating mechanism then a new technique must be found to solve these constraints.

The original motivation for fermionic T-duality was to understand the dual superconformal invariance found in maximally supersymmetric Yang-Mills theory. Similarly, it is hoped that there will an understanding of the dual superconformal symmetry of ABJM \cite{bargheer, huang} using fermionic T-duality in type IIA theory. The string theory dual to ABJM \cite{ABJM} theory is type IIA string theory on AdS$_4 \times \mathbb{CP}^3,$ and there has been work on trying to understand the self-duality of the AdS$_4 \times \mathbb{CP}^3$ background under a combination of T-duality and fermionic T-duality \cite{adam, grassi, ado2}. In \cite{ado2}, fermionic T-duality transformations on the partially $\kappa$--gauge fixed Green-Schwarz action is considered and found to be singular. However, the partially $\kappa$-gauge fixed action for the $AdS_{4} \times \mathbb{C}P^{3}$ sigma model is not well-defined for all string configurations. It is not clear in \cite{ado2} whether the singularity arises for this reason or not. The transformation rules for the type IIA supergravity fields derived in this paper can be used to perform the transformation from the target space point of view. Indeed this has recently been done  by Bakhmatov \cite{bakhmatov}. The results of this paper are consistent with the singularity found in \cite{ado2}. In \cite{bakhmatov} the transformation is done solely in supergravity, and hence the work suggests that the singularity found in \cite{ado2} does not have its source in the sigma model. It is an intriguing problem to find out the origin of this singularity at the supergravity level.

Finally, in the transformation rules for type IIA supergravity besides the conditions which have analogues in the type IIB supergravity transformations we also found that 
\begin{equation*}
\bar{\epsilon}_{I} \gamma_{11} \eta_{J} M_{IJ}=0
\end{equation*}
must hold. This condition may be physically interpreted as maintaining a zero Romans mass \cite{massiveIIA}, for the Romans mass can be thought of as a constant 0-form field strength \cite{D8}. This suggests that the fermionic symmetry that we have constructed for type IIA supergravity can be extended to massive type IIA supergravity. We will report on this problem in a future paper.

Another problem that we would like to address in the future is the complexity of the fermionic T-duality transformations, for in string theory the transformations cannot be made real. Understanding the physical interpretation of the complexity may reveal important, hitherto unknown, aspects of string theory.

\section*{Acknowledgements}
We would like to thank D. Berman, J. Gutowski, A. Maharana, J. Santos and M. Wolf for discussions. HG is supported by the STFC.

\appendix
\section{Conventions}
\label{conv}

Below we summarise the conventions used in this work.

The metric signature is $(- +\dots+).$

The permutation symbol is totally antisymmetric and its sign is defined by $$ \epsilon_{01\dots9}=1.$$

For a $p-$form $A$ and $q-$form $B$ 
$$\left( \d A \right)_{a_1 \dots a_{p+1}} = (p+1) \partial_{[a_1} A_{a_2 \dots a_{p+1}]},$$
$$\left( A \wedge B \right)_{a_1 \dots a_p b_1 \dots b_q}= \frac{(p+q)!}{p!q!} A_{[a_1 \dots a_p } B_{b_1 \dots b_q]},$$
$$ (\star A)_{a_1 \dots a_{d-p}}=\frac{1}{p!} \epsilon_{a_1 \dots a_{d-p}}^{\qquad \quad b_1 \dots b_{p}} A_{b_1 \dots b_{p}}.$$

The chirality matrix is $$ \gamma_{11}=- \gamma_0 \gamma_1 \dots \gamma_9.$$

\section{Gamma matrix identity}
\label{gammadual}
Let $\{\gamma_{a}\}$ be a matrix representation of the $10$-dimensional Clifford algebra. Since $\gamma_{c_1 \dots c_{10}} = -\epsilon_{c_1 \dots c_{10}} \gamma_{11},$
 \begin{align*}
&\epsilon_{a_1 \dots a_m b_1 \dots b_{10-m}} \gamma_{11} \gamma^{b_1 \dots b_{10-m}} \\
= & \epsilon_{a_1 \dots a_m b_1 \dots b_{10-m}} \left( \frac{1}{{10}!} \epsilon^{c_1 \dots c_{10}} \gamma_{c_1 \dots c_{10}} \right)  \gamma^{b_1 \dots b_{10-m}} \\
= &  \frac{1}{{10}!} \epsilon_{a_1 \dots a_m b_1 \dots b_{10-m}} \epsilon^{c_1 \dots c_{10}}  \sum_{k=0}^{10-m} c_{10(10-m)}^k \gamma_{[c_1 \dots c_{10-k}}^{\qquad \quad \;\; [b_1 \dots b_{10-m-k}} \delta_{c_{10-k+1}}^{b_{10-m-k+1}} \dots \delta_{c_{10}]}^{b_{10-m}]},
 \end{align*}
using identity \eqref{gammaidentity}. Now, contracting the Kronecker delta functions with $\epsilon^{c_1 \dots c_{10}},$ the expression becomes
\begin{align*}
& \frac{1}{{10}!} \epsilon_{a_1 \dots a_m b_1 \dots b_{10-m}} \sum_{k=0}^{{10}-m} c_{{10}({10}-m)}^k \epsilon^{c_1 \dots c_{{10}-k} b_{{10}-m-k+1} \dots b_{{10}-m}} \gamma_{c_1 \dots c_{{10}-k}}^{\qquad \quad \;\; b_1 \dots b_{{10}-m-k}}\\
=&  \sum_{k=0}^{{10}-m} \frac{c_{{10}({10}-m)}^k}{10!} \left( - ({10}-k)!k! \delta_{[a_1}^{[c_1} \dots \delta_{a_m}^{c_m} \delta_{b_1}^{c_{m+1}} \dots \delta_{b_{{10}-m-k}]}^{c_{{10}-k}]} \right) \gamma_{c_1 \dots c_{{10}-k}}^{\qquad \quad \;\; b_1 \dots b_{{10}-m-k}} \\
=& - \sum_{k=0}^{{10}-m} \frac{c_{{10}({10}-m)}^k}{10!} ({10}-k)!k!  \gamma_{a_1\dots a_m b_1 \dots b_{{10}-m-k}}^{\qquad \qquad \qquad \quad b_1 \dots b_{{10}-m-k}},
\end{align*}
but $$\gamma_{a_1\dots a_m b_1 \dots b_{{10}-m-k}}^{\qquad \qquad \qquad \quad b_1 \dots b_{{10}-m-k}}=0,$$
unless $k={10}-m.$ Therefore, 
\begin{align*}
\epsilon_{a_1 \dots a_m b_1 \dots b_{{10}-m}} \gamma_{11} \gamma^{b_1 \dots b_{{10}-m}} & = -\frac{1}{{10}!} c_{{10}({10}-m)}^{{10}-m} m!({10}-m)!  \gamma_{a_1\dots a_m } \\
&= - (-1)^{({10}-m)^2 + \frac{1}{2} ({10}-m)({10}-m+1)} ({10}-m)!  \gamma_{a_1\dots a_m }.
\end{align*}
Hence,
\begin{equation*}
\gamma_{a_1 \dots a_m}= -\frac{(-1)^{\frac{1}{2}({10}-m)({10}-m+1)}}{({10}-m)!} \epsilon_{a_1 \dots a_m b_1 \dots b_{{10}-m}} \gamma^{b_1 \dots b_{{10}-m}} \gamma_{11}. 
\end{equation*}

\bibliography{symm}
\bibliographystyle{utphys}
\end{document}